\documentclass[pra,letterpaper,aps,10pt,superscriptaddress,twocolumn,floatfix,showpacs]{revtex4-2}
\usepackage{graphicx}
\usepackage{amsmath}
\usepackage{amsfonts}
\usepackage{float}
\usepackage{amssymb}
\usepackage{epsfig}
\usepackage{epstopdf}
\DeclareGraphicsExtensions{.pdf,.eps,.png,.jpg,.mps}
\usepackage[pdftex]{color}
\usepackage{amsmath,graphicx,amssymb,braket,xcolor,subfigure,upgreek}
\usepackage[colorlinks, linkcolor=blue, citecolor=blue, urlcolor=blue, breaklinks=true]{hyperref}
\usepackage{microtype}
\usepackage{bbm}
\usepackage{color}
\usepackage{dsfont}

\begin{document}

\title{Prospects of phase-adaptive cooling of levitated magnetic particles in a hollow-core photonic-crystal fiber}
\author{P. Kumar}
\email{pardeep.kumar@mpl.mpg.de}
\affiliation{Max Planck Institute for the Science of Light, Staudtstra{\ss}e 2,
D-91058 Erlangen, Germany}
\author{F. G. Jimenez}
\affiliation{Pontificia Universidad Católica del Perú, Av. Universitaria 1801, San Miguel 15088, Peru}
\author{S. Chakraborty}
\affiliation{Department of Physics, Friedrich-Alexander-Universit\"{a}t Erlangen-N\"urnberg, Staudtstra{\ss}e 7,
	D-91058 Erlangen, Germany}
\affiliation{Max Planck Institute for the Science of Light, Staudtstra{\ss}e 2,
	D-91058 Erlangen, Germany}
\author{G. K. L. Wong}
\affiliation{Max Planck Institute for the Science of Light, Staudtstra{\ss}e 2,
D-91058 Erlangen, Germany}
\author{N. Y. Joly}
\affiliation{Department of Physics, Friedrich-Alexander-Universit\"{a}t Erlangen-N\"urnberg, Staudtstra{\ss}e 7,
	D-91058 Erlangen, Germany}
\affiliation{Max Planck Institute for the Science of Light, Staudtstra{\ss}e 2,
D-91058 Erlangen, Germany}
\author{C. Genes}
\affiliation{TU Darmstadt, Institute for Applied Physics, Hochschulstra{\ss}e 4A, D-64289 Darmstadt, Germany}
\affiliation{Max Planck Institute for the Science of Light, Staudtstra{\ss}e 2,
D-91058 Erlangen, Germany}
\affiliation{Department of Physics, Friedrich-Alexander-Universit\"{a}t Erlangen-N\"urnberg, Staudtstra{\ss}e 7,
D-91058 Erlangen, Germany}
\date{\today}

\begin{abstract}
We analyze the feasibility of cooling of classical motion of a micro- to nano-sized magnetic particle, levitated inside a hollow-core photonic crystal fiber. The cooling action is implemented by means of controlling the relative phase between counter-propagating fiber guided waves. Direct imaging of the particle’s position allows dynamic phase adjustments that produce a Stokes-type cooling force. We provide estimates of cooling efficiency and final achievable temperature, taking into account thermal and detection noise sources. Our results bring forward an important step towards using trapped micro-magnets in sensing, testing the fundamental physics and preparing the quantum states of magnetization.
\end{abstract}


\maketitle

%%%%%%%%%%%%%%%%%%%%%%%%%%%%%%%%%%%%%%%%%%
%%%%%%%%%%%%%%%%%%%%%%%%%%%%%%%%%%%%%%%%%%
\section{Introduction}
%%%%%%%%%%%%%%%%%%%%%%%%%%%%%%%%%%%%%%%%%%
%%%%%%%%%%%%%%%%%%%%%%%%%%%%%%%%%%%%%%%%%%
\label{sec1}
Mechanical resonators are good actuators for displacement~\cite{anetsberger2009displacement}, force~\cite{fogliano2021force}, acceleration~\cite{krause2012acceleration} or mass sensing~\cite{sansa2020mass} applications. An essential requirement of the nano-mechanical devices is to eliminate the heating and decoherence caused by clamping. In this context, optical levitation of mesoscopic particles, set an exceptional platform to achieve unprecedented sensing limits. The absence of clamping losses, a unique feature of such systems, is achieved by using a tightly-focused optical beam to trap micro and nanoparticles in free space \cite{BallesteroScience2021}. These optical traps are well-suited for advancing room-temperature quantum optomechanics \cite{BarkerPRA2010}, investigating non-equilibrium physics \cite{GieselerNatNano2014} and testing the fluctuation-dissipation theorem \cite{GieselerEntropy2018}.\\ 

Although free space optical tweezers possess tremendous sensing capabilities, their performance is limited by diffraction and issues in the optical manipulation of trapped particles. These limitations can be addressed by trapping micro-and nanoparticles in hollow-core photonic crystal fiber (HC-PCF). This configuration offers unique advantages, including diffraction free trapping and possibility to tune the environment surrounding the particle by introducing gas, vacuum or liquid  inside the hollow-core \cite{SharmaOptL2021,ZeltnerAPL2016}. Another key advantage of optical trapping in HC-PCF is the ability to guide  trapped particles over distances several orders of magnitude larger than the Rayleigh length of the trapping light \cite{BenabidOptExp2002,RennPRL1995,BykovNatPhys2015,BykovLSA2018}. This is achieved by means of an optical conveyor belt formed from the standing wave optical trap created by two counter-propagating beams. The particles can then be transported or even halted at a particular position in the fiber by changing  the relative phase \cite{SchmidtOptExp2013} between counter-propagating modes.\\

In above described configurations, a prerequisite for sensing is the control of mechanical oscillator's motion to a level where fluctuations, either classical, such as stemming from thermal environments or quantum, are almost completely frozen. In the optomechanical realm~\cite{aspelmeyer2014rmp}, cooling of isolated mechanical resonances has been achieved both via cavity self-cooling~\cite{aspelmeyer2019coherent,uros2020cooling,windey2019,gonzalez2019cavitycooling}, feedback in the form of cold-damping~\cite{li2011millikelvin,millen2020review,rossi2018measurement,tebbenjohanns2021quantum,magrini2021real,whittle2021approaching} or parametric modulation~\cite{villanueva2011a,gieseler2012subkelvin,gieseler2013thermal,rodenburg2016quantum,sames2018continuous,ghosh2023theory}.\\

%%%%%%%%%%%%%%%%%%%%%%%%%%%%%%%%%%%%%%%%%%%%%%%
\begin{figure}[t]
\includegraphics[width=1.0\columnwidth]{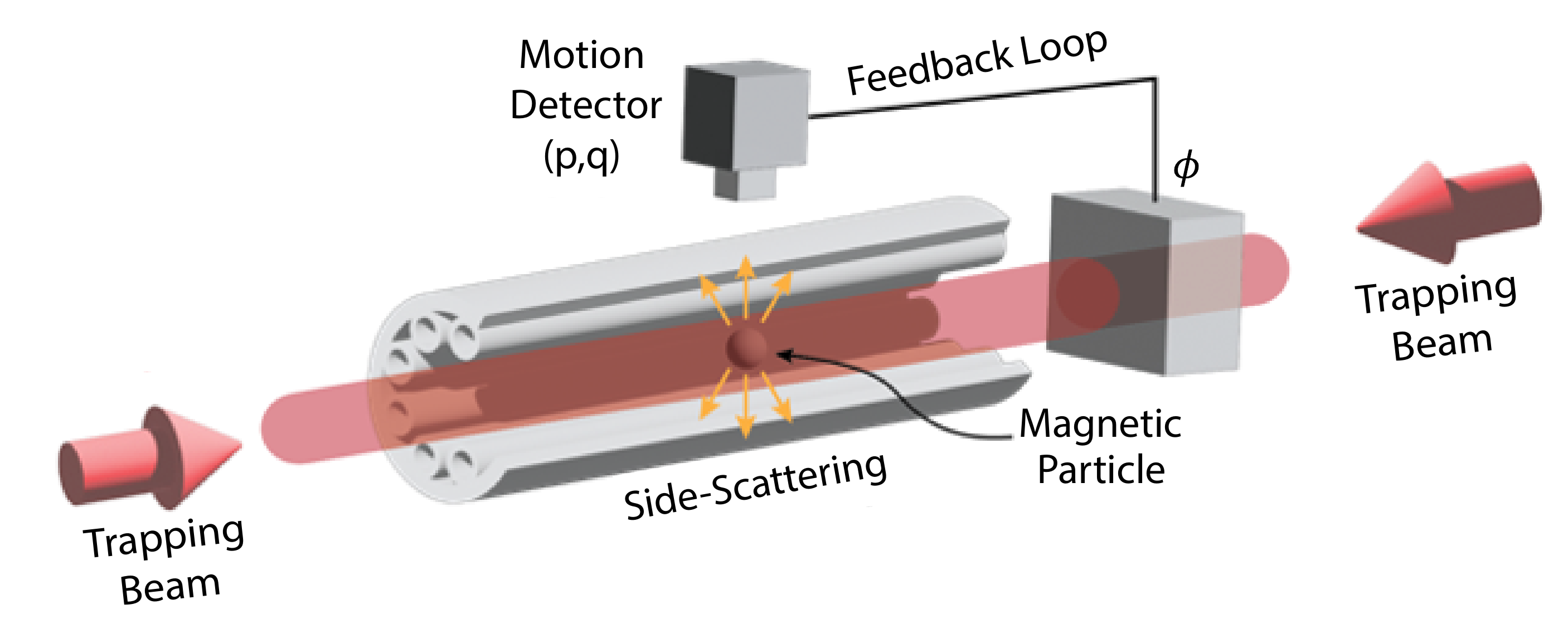}
\caption {A magnetic micro-sphere is trapped in the potential minimum of a standing wave inside a hollow-core optical photonic crystal fiber. The standing wave minima can be shifted in the $y$-direction of the fiber by the modification of the phase $\phi(t)$ of the right-propagating guided control laser. Detection of the particle's position $y_\text{det}(t)$ gives an estimate $y_\text{det}$ of the true value $y(t)$, with an additional noise term $\eta W(t)$ modeled as a Wiener process $W(t)$ multiplied by the efficiency of the imaging scheme (poorer resolution translates into a larger $\eta$). Phase adaptive feedback results from the monitoring of the detected quadrature $y_\text{det}$ and the subsequent adjustment of $\phi$.}
\label{fig1}
\end{figure}
%%%%%%%%%%%%%%%%%%%%%%%
\indent    Magnetic micro-particles composed of yttrium iron garnet and yttrium orthoferitte have been recently trapped inside photonic crystal fibers \cite{chakraborty2024arXiv}.  Owing to the large magnetization \cite{ZhangSciAdv2016,CherepanovPhysRep1993} of the trapped particle, such a system holds the promise of providing excellent sensors for stray magnetic fields \cite{ZareRameshtiPhysRep2022}. However, at moderate temperatures, their occupancy is at the level of $10^{10}$ phonons, as their trapping frequency $\Omega$ lies typically in the kilohertz regime. In addition, the gas pressure which is required to keep the particle stably trapped, around mbar, gives rise to a damping rate at the level of $300$~Hz meaning that the mechanical quality factor is very low. However, further reduction of the pressure, desirable in order to increase the mechanical quality factor, causes the particle to escape from the trap due to increase in its bulk temperature and insufficient damping at low pressure \cite{ricciThesis2019}. \\
\indent We propose here an alternative route, where external optical damping is switched on during the pressure reduction stage and switched off in the next stage. This ensures the stability of the system and allows for the reach of very low pressures, where the particle is isolated from the thermal environment. The feedback loop involves a phase-adaptive mechanism \cite{ghosh2023theory}, providing an effective Stokes-type damping force \cite{genes2008ground}. The feedback force comes from the continuous adjustment of the relative phase $\phi(t)$ between the two counter-propagating trapping lasers (see Fig.~\ref{fig1}). This consequently allows for a controllable shifting of the potential minimum and thus the particle's equilibrium position. In a first step, one detects light scattered to the side of the fiber and feeds the information into a motion controller. The deduced quantity is the particle's dimensionless position $q_\text{det}(t)=q(t)+\eta W^{\text{det}}(t)$, which contains the exact value of the position quadrature $q(t)$, to which one adds inherent detection noise of amplitude $\eta$ and modelled as a Wiener process $W^{\text{det}}(t)$ \cite{Jacobs2014}. The detected signal is then used to estimate $\phi(t)=(\mathcal{C}/\Omega)[q_\text{det}(t)-q_\text{det}(t-\tau)]/\tau$, where $\mathcal{C}$ and $\tau$ are adjustable parameters. Subsequently, the phase deduced is used to control the right propagating laser, leading to the desired optical damping effect. Notably, Grass et. al. \cite{GrassAPL2016} have demonstrated the trapping and cooling of a dielectric silica particle inside HC-PCF. The feedback cooling action in \cite{GrassAPL2016} is achieved by using an additional laser and modulating the power of the optical trap. However, phase adaptation in our technique is fundamentally different and requires only the adjustment of the relative phase, unlike power, between counter-propagating beams. One of the main advantages of phase actuation is the absence of extraneous heating that may arise from the direct intensity modulation. Another advantage of phase actuation is the controlled transportation of the cold particle from one place to another. This is not possible by simply changing the power \cite{GrassAPL2016}. \\

\indent The paper is organized as follows. In Sec. \ref{sec2}, we introduce the model and equations of motion for a classically trapped particle subject to stochastic noise stemming from the thermal environment and the dynamical adjustment of the phase. We then provide analytical solutions for the optical damping rate as well as for the final achievable occupancy in Sec. \ref{sec3}. We model the backaction noise stemming from the action of the feedback loop. In Sec. \ref{sec4}, we provide experimental description for the system under consideration, an in-depth discussion on the tunability of the involved parameters to freeze thermal fluctuations from the system and validation of the results by accompanying numerical simulations. Finally, the concluding remarks are presented in Sec. \ref{sec5}\\

%%%%%%%%%%%%%%%%%%%%%%%%%%%%%%%%%%%%%%%%%%
%%%%%%%%%%%%%%%%%%%%%%%%%%%%%%%%%%%%%%%%%%
\section{Model and equations}
\begin{figure}[t]
	\includegraphics[width=0.99\columnwidth]{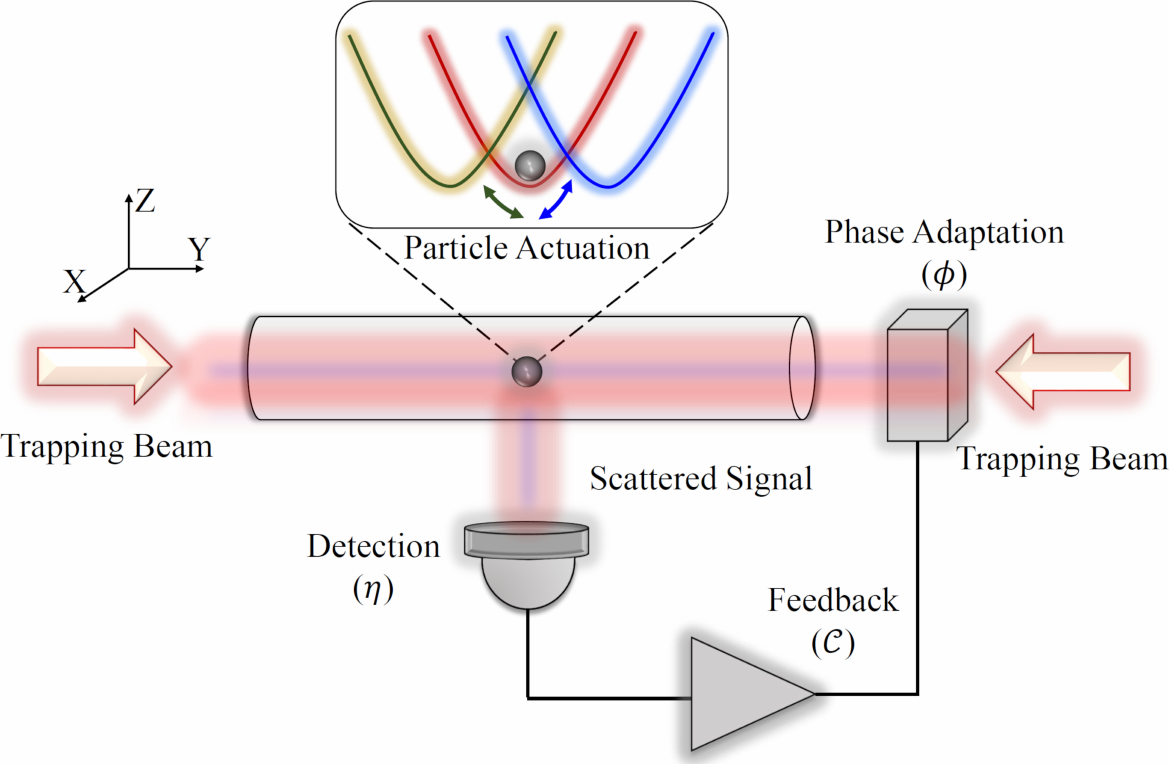}
	\caption {Detailed scheme showing the important parameters involved in the feedback loop, such as the imaging factor $\eta$ and the feedback amplification factor $\mathcal{C}$. The inset shows the shifting of the trap minimum (red harmonic potential) from the equilibrium position to the left (green) and right (blue) via the control of the right pump phase $\phi$. This control knob allows for the emergence of a Stokes-like optical damping force, acting to cool the particle's motion.}
	\label{fig2}
\end{figure}

%%%%%%%%%%%%%%%%%%%%%%%%%%%%%%%%%%%%%%%%%%
%%%%%%%%%%%%%%%%%%%%%%%%%%%%%%%%%%%%%%%%%%
\label{sec2}
Let us assume that a magnetic particle of mass $m$ is trapped in the standing wave formed by two counter-propagating waves with frequency $\omega_\ell$ and wave-vector $\pm k_\ell$ along the $y$-direction. The plus component of the electric field amplitude can be expressed as a superposition of $\mathcal{E}_{0}e^{ik_\ell y}e^{-i \omega_\ell t}$ and $\mathcal{E}_{0}e^{-ik_\ell y} e^{-i \omega_\ell t} e^{-2i\phi(t)}$ counterpropagating waves,
where $\mathcal{E}_{0}$ and $\phi(t)$ are the amplitude and the relative phase of the light fields. In a frame rotating at $\omega_\ell$, the total electric field is then $\mathcal{E}(y,t)=2\mathcal{E}_{0}\cos[k_\ell y+\phi(t)]$. This leads to a position and time dependent potential energy $-\alpha' \mathcal{E}(y,t)^2/4$ for the particle, where $\alpha'$ represents the real part of the particle's polarisability. For small displacement from the center of the standing wave trap, our theoretical model assumes a dipole-like gradient force on the particle's center of mass (see Appendix \ref{AppendixC1} for details). This allows in turn for the computation of the gradient optical force, in the $y$-direction, to be obtained as
\begin{align}
	F_{y}&=\alpha'|\mathcal{E}_{0}|^{2}\frac{\partial}{\partial y}\Big[\cos(k_\ell y+\phi(t)\Big]^2\;.
\end{align}
 Assuming that the particle is trapped in the minimum of the optical potential, it is easy to decompose the force into the following two contributions
\begin{align} F_{y}&=-\Big[2k^2_\ell\alpha'|\mathcal{E}_{0}|^{2}\Big]y-\Big[2k_\ell\alpha'|\mathcal{E}_{0}|^{2}\Big]\phi(t).
\end{align}
The first term in above equation is the restoring force, indicating a spring (or trapping) constant $\kappa_\text{trap}=2k_\ell^{2}\alpha'|\mathcal{E}_{0}|^{2}=m\Omega^{2}$. This way we set the value of the optical trap frequency $\Omega$ and note that it can be adjusted solely by tuning the laser power, as the mass $m$ and the real part of the polarizibility $\alpha'$ are fixed values for a given  material used. The second term is an additional time-dependent force, arising due to the relative phase difference between the counter-propagating fields and expressed in a simpler form as $F_\text{add}=-(\kappa_\text{trap}/{k_\ell})\phi(t)$. By adjusting the right pump's phase $\phi$, this force acts as an actuator by shifting the trap minimum from its equilibrium position, in such a way to counteract its motion when desired, as shown in the inset of Fig. \ref{fig2}.\\
%%%%%%%%%%%%%%%%%%%%%%%%%%%%%%%%%%%%%%%%%%
%%%%%%%%%%%%%%%%%%%%%%%%%%%%%%%%%%%%%%%%%%
%%%%%%%%%%%%%%%%%%%%%%%%%%%%%%%%%%%%%%%%%%
%%%%%%%%%%%%%%%%%%%%%%%%%%%%%%%%%%%%%%%%%%
\indent We can now proceed in writing the difference equations for the oscillator subject to the additional force stemming from the dynamical adjustment of the phase $\phi(t)$ and to thermal noise as
\begin{subequations}
\begin{align}
dq &= \Omega pdt\;,\\
dp &=-\gamma pdt-\Omega qdt-\beta \phi(t) dt+\sqrt{2\gamma n_{\mathrm{th}}}dW^{\text{th}}(t)\;.
\end{align}
\label{EqsMotion}
\end{subequations}
In doing so we have reduced the description to dimensionless quadratures $y=y_\text{zpm}q$ and $p=p_\text{zpm}p$ where the zero-point motion for the position is $y_\text{zpm}=\sqrt{\hbar/(m\Omega)}$ (the extent of the wavefunction at zero temperature) and for the momentum $p_\text{zpm}=\sqrt{\hbar m\Omega}$. Here, $\hbar$ is the reduced Planck's constant. Also, we introduced the notation $\beta=\kappa_\text{trap}/({k_\ell}p_{\mathrm{zpm}})$, with dimensions of frequency. The Wiener increment for the thermal environment has zero average and only surviving correlations $\braket{dW^{\text{th}}(t)dW^{\text{th}}(t)}=dt$ at equal times (all correlations at different times vanish). The thermal occupancy corresponding to a bath of temperature $T_{\mathrm{th}}$ (assumed large) is written as $n_{\mathrm{th}}\approx k_B $T$_{\mathrm{th}}/(\hbar\Omega) $, where $k_B$ is the Boltzmann constant. For example, assuming $\Omega/2\pi=34.5~$kHz and $T_{\mathrm{th}}=300~$K, one obtains $n_{\mathrm{th}}\sim10^{8}$, very far from the quantum regime. From Eqs.~(\ref{EqsMotion}), it can be easily verified that in the absence of additional actuation force, the system achieves equilibrium at a rate $\gamma$ with its thermal environment at temperature $T_{\mathrm{th}}$ (see Appendix \ref{AppendixF} for details).\\
\indent We provide now a quasi-derivative feedback cooling procedure, where the phase of the control laser is assumed to be
\begin{equation}
\phi(t)=\frac{\mathcal{C}}{\Omega} \times \frac{q_\text{det}(t)-q_\text{det}(t-\tau)}{\tau}\;.\label{EqPhase}
\end{equation}
Here $\mathcal{C}$ is the feedback amplification factor: larger values lead to faster cooling but also increased detection noise. The time interval $\tau$ is the interval on which the quasi-derivative operation is performed. A direct cold-damping feedback is obtained by setting $\tau=dt$ and letting $dt\rightarrow 0$. The detection noise is modelled by the inclusion of a Wiener process to the true value of the position and multiplied by an amplitude $\eta$ such that at any time $q_\text{det}(t)=q(t)+\eta W^{\text{det}}(t)$. This allows to estimate the best strategy in terms of the choice of $\tau$ and in terms of the optimal magnitudes of $\mathcal{C}$ and $\eta$. We will address this procedure in Sec.~\ref{sec3} and estimate both the cooling rate as well as final occupancy encompassing both thermal and measurement noise contributions.\\

%%%%%%%%%%%%%%%%%%%%%%%%%%%%%%%%%%%%%%%%%%
%%%%%%%%%%%%%%%%%%%%%%%%%%%%%%%%%%%%%%%%%%
\section{Theory of quasi-derivative cooling}
%%%%%%%%%%%%%%%%%%%%%%%%%%%%%%%%%%%%%%%%%%%%%%%%%%%%%%%%%%%%%%%%%%%%%%
%%%%%%%%%%%%%%%%%%%%%%%%%%%%%%%%%%%%%%%%%%%%%%%%%%%%%%%%%%%%%%%%%%%%%%
\label{sec3}
We will separate our analysis in two parts by first deducing an analytical expression for the optical feedback cooling rate and afterwards estimating the added measurement noise and the corresponding final achievable occupancy. In a first step we solely consider deterministic motion (no thermal or measurement noise) and deduce the feedback effect in adding an extra cooling rate $\bar{\Gamma}$ to $\gamma$. In the second step, we estimate the contributions of both (uncorrelated) noise sources, detection and thermal, and deduce the effective occupancy incorporating both effects.\\

%%%%%%%%%%%%%%%%%%%%%%%%%%%%%%%%%%%%%%%%%%
%%%%%%%%%%%%%%%%%%%%%%%%%%%%%%%%%%%%%%%%%%
\subsection{Cooling rate}
To obtain the modified cooling rate, we take into account of the deterministic part of Eqs.~\eqref{EqsMotion} and neglect the influence of noise. This allows us to write the following set of coupled differential equations
\begin{subequations}
\begin{align}
\frac{dq}{dt} &= \Omega p\;,\\
\frac{dp}{dt} &=-\gamma p-\Omega q-\beta \phi(t)\;.
\end{align}
\label{EqsMotionNoNoise}
\end{subequations}
A formal integration of the first equation for the displacement quadrature yields $q(t)=q(t-\tau)+\Omega\int_{t-\tau}^{t}p(t')dt'$. Assuming that the effect of the feedback is not too strong in the interval from $t-\tau$ to $t$ (which is valid when restricting values of $\tau$ of the order of the mechanical period $T=2\pi/\Omega$, or smaller), one can assume, in first order, that only natural evolution at frequency $\Omega$ and damping rate $\gamma$ occurs in this interval $t-\tau$ to $t$. This allows one to deterministically connect $p(t')$ with $p(t)$ and $q(t)$. After performing the integration of decaying, oscillating terms (see Appendix~\ref{AppendixA}), one sees a modification of the effective decay rate from the purely thermally induced $\gamma$ to the optically modified one $\gamma+\bar{\Gamma}$ (by identifying the terms proportional to $p(t)$) where
\begin{equation}
\bar{\Gamma}=\beta \mathcal{C}\Big[e^{\gamma\tau/2}~\frac{\sinh{\tilde{\Omega} \tau}}{\tilde{\Omega}\tau}\Big].
\end{equation}
Here we introduce $\tilde{\Omega}=\sqrt{\gamma^{2}-4\Omega^{2}}$. For $\tau\rightarrow 0$, we recover an expected cold-damping type of result with a derivative feedback showing an optimal value $\Gamma=\beta \mathcal{C}$.

In addition to damping, the mechanical resonance frequency is also modified via an optical spring effect, which we obtain from the identification of the extra factors in front of $q(t)$ leading to
\begin{equation}
	\bar \Omega =\Omega-\beta \mathcal{C}\left[\frac{e^{\gamma \tau/2}\cosh{(\tilde{\Omega}\tau})-1}{\Omega \tau}\right]+\frac{\gamma\bar{\Gamma}}{2\Omega}\;.
\end{equation}

In the limit of small $\tau$, the optical spring effect cancels out $\bar \Omega=\Omega$. However, for finite $\Omega \tau$, this indicates an additional tuning knob for reducing the effective occupancy, by increasing the oscillator frequency.

%%%%%%%%%%%%%%%%%%%%%%%%%%%%%%%%%%%%%%%%%%
%%%%%%%%%%%%%%%%%%%%%%%%%%%%%%%%%%%%%%%%%%

\subsection{Final occupancy}
\begin{figure}[t]
	\includegraphics[width=0.95\columnwidth]{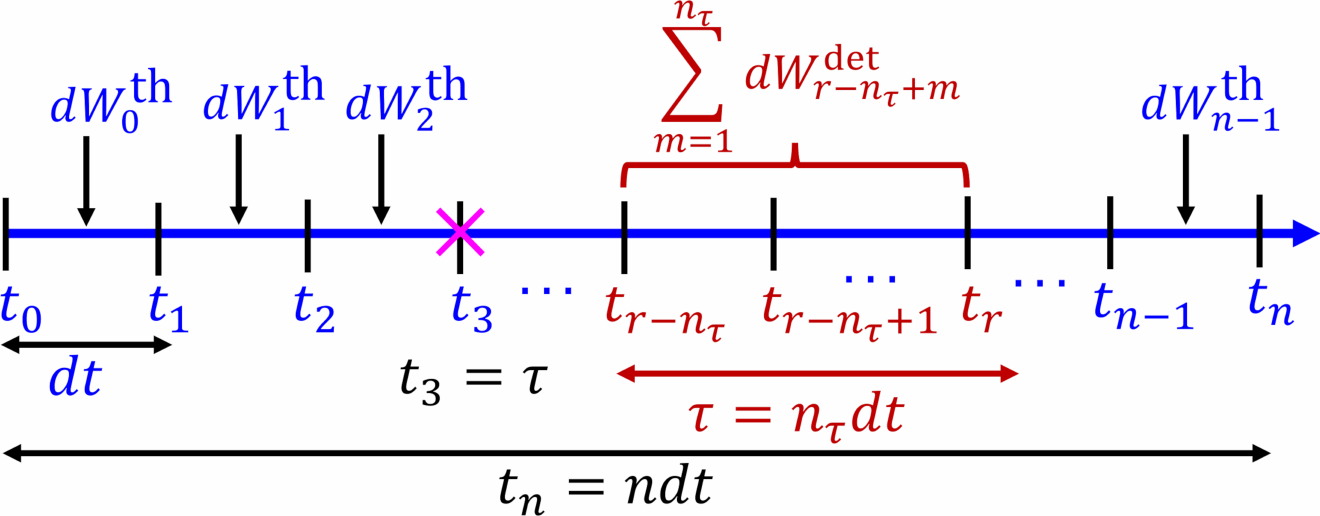}
	\caption {Schematic depicting the time steps in which thermal and measurement back-action noise are included during the time evolution. The time interval $[t_{0},t_{n}]$ is chopped into $n$ infinitesimal intervals of length $dt=t_{n}/n$. The thermal noise increments, $dW^{\mathrm{th}}_{j}~(j=0,1\cdots n-1)$, are added in each time-step. The phase adaptive feedback starts at time $\tau$ shown by magenta cross and is discretized into $n_{\tau}$ steps of duration $dt$. The measurement noise is added as sum over Wiener increments for each evolution step, once the feedback starts acting. One of such an interval from $t_{r-n_{\tau}}$ to $t_{r}$ is shown in red during which the Wiener increments $\sum_{m=1}^{n_{\tau}}dW_{r-n_{\tau}+m}^{\mathrm{det}}$ are included.}
	\label{fig3}
\end{figure}

%%%%%%%%%%%%%%%%%%%%%%%%%%%%%%%%%%%%%%%%%%%%%%%%%%%%%%%%%%%%%%%%%%%%%%
%%%%%%%%%%%%%%%%%%%%%%%%%%%%%%%%%%%%%%%%%%%%%%%%%%%%%%%%%%%%%%%%%%%%%%
Let us now analyze how measurement noise enters the equation of motion Eq.~(\ref{EqsMotion}b) via the term $-\beta \phi(t) dt$. This can be written in the difference equations form as
\begin{equation}
\mathcal{N}^\text{det}(t) = -\frac{\eta \beta \mathcal{C}}{\Omega} \left[\frac{W^{\text{det}}(t)-W^{\text{det}}(t-\tau)}{\tau}\right]dt.
\end{equation}
and to be added to the standard thermal noise $\sqrt{2\gamma n_{\mathrm{th}}}dW^{\text{th}}(t)$. By expressing the Wiener process entering the detection as a discrete sum over Wiener increments [see Fig. \ref{fig3}] with infinitesimal increment of time $dt$ (such that $t=ndt$ and $\tau=n_\tau dt$), we can rewrite
\begin{equation}
\mathcal{N}^\text{det}(t) = -\frac{\eta \beta \mathcal{C}dt}{\Omega \tau}\sum_{m=1}^{n_\tau}dW^{\text{det}}_{n-n_\tau+m}.
\end{equation}
%%%%%%%%%%%%%%%%%%%%%%%%%%%%%%%%%%%%%%%%%%%%%%%%%%%%%%%%%%%%%%%%%%%%%%%%%%%%%%%%%%%%%%%%%%%%%
%%%%%%%%%%%%%%%%%%%%%%%%%%%%%%%%%%%%%%%%%%%%%%%%%%%%%%%%%%%%%%%%%%%%%%
%%%%%%%%%%%%%%%%%%%%%%%%%%%%%%%%%%%%%%%%%%%%%%%%%%%%%%%%%%%%%%%%%%%%%%
%%%%%%%%%%%%%%%%%%%%%%%%%%%%%%%%%%%%%%%%%%%%%%%%%%%%%%%%%%%%%%%%%%%%%%
%%%%%%%%%%%%%%%%%%%%%%%
In order to estimate the contribution of the two types of noises to the steady state of the system, we follow the derivation shown in Ref.~\cite{ghosh2023theory} and detailed in Appendix~\ref{AppendixB}. The steady state solution for the momentum is obtained by taking the large $n$ limit and reads

\begin{equation}
	p_n=\frac{1}{\bar{\gamma}_{+}-\bar{\gamma}_{-}}\sum_{j=0}^{n-1}\Big(\mathcal{N}_{j}^{det}+\sqrt{2\gamma n_\text{th}}dW^{th}_{j}\Big)\Big(\bar{\gamma}_{+}\bar{\lambda}_{+}^{j}-\bar{\gamma}_{-}\bar{\lambda}_{+}^{j}\Big)\;,
\end{equation}
where the coefficients are obtained from the eigenvalues, $\bar{\lambda}_{\pm}=1-\bar{\gamma}_{\pm}dt/2$, of the evolution matrix. Here
\begin{equation}
\bar{\gamma}_{\pm}=\bar{\gamma}\pm\sqrt{\bar{\gamma}^{2}-4\bar{\Omega}^{2}},
\end{equation}
contain the renormalized damping rate $\bar \gamma=\gamma+\bar{\Gamma}$ and frequency $\bar \Omega$. The aim is to compute the variance $\braket{p_n^2}$ in steady state. As the two noises are uncorrelated one can separate two contributions $\braket{p_n^2}=\braket{p_n^2}_{\text{th}}+\braket{p_n^2}_{\text{det}}$ and proceed by estimating each of them independently.

In a first step, we use the thermal noise correlations $\braket{dW_j^{\text{th}}dW_{j'}^{\text{th}}}=dt\delta_{jj'}$ and make use of the following identities $\lim_{n\rightarrow\infty}\sum_{j=0}^{n-1} \bar{\lambda}_{\pm}^{2j}=1/(\bar{\gamma}_{\pm} dt)$, and $\lim_{n\rightarrow\infty}\sum_{j=0}^{n-1} \bar{\lambda}_{+}^{j}\bar{\lambda}_{-}^{j}=2/[(\bar{\gamma}_{+}+\bar{\gamma}_{-}) dt]$ (for details see Appendix~\ref{AppendixB}). This gives the correct estimate of the thermal final occupancy of the oscillator undergoing extra damping at rate $\bar{\Gamma}$ and modified frequency $\bar{\Omega}$:
\begin{equation}
\braket{p_n^2}_{\text{th}}=\frac{\gamma n_\text{th}}{\bar{\gamma}}\Big(\frac{\Omega}{\bar{\Omega}}\Big).
\end{equation}
As expected, the initial large occupancy is reduced to the ratio of the thermalization rate $\gamma n_{\mathrm{th}}$ divided by the effective total damping rate $\bar \gamma$ but with a modification resulting from optical spring effect \cite{corbittPRL2007}. In conventional feedback techniques, the final thermal occupancy can be reduced by increasing the damping rate. On top of this, enhanced resonant frequency provides an additional cooling \cite{corbittPRL2007_2}. The combined result of these two effects can yield very low thermal occupancy.\\
\indent Let us now compute the steady state correlations of the detection noise $\braket{p_n^2}_\text{det}$. First we introduce $\mathcal{A}=-\eta \beta \mathcal{C}dt/(\Omega \tau)$ and rewrite $\mathcal{N}_j=\mathcal{A} \sum_{m=1}^{n_\tau}dW^{\text{det}}_{n-n_\tau+m}$. Taking into account that the correlations $\braket{dW^{\text{det}}_{j-n_\tau+m}dW^{\text{det}}_{j'-n_\tau+m'}}$ will impose the constraint $j+m=j'+m'$ leads us to the following expression for the detection induced noise in the momentum quadrature
\begin{equation} \braket{p_n^2}_\text{det}=\frac{\mathcal{A}^{2}dt}{(\bar{\gamma}_{+}-\bar{\gamma}_{-})^{2}}\sum_{j=0}^{n-1}\sum_{m,m'=0}^{n_{\tau}}\Big[f_{j}-g_{j}^{m-m'}\Big]\;,
\end{equation}
where \begin{equation}
f_{j}=\bar{\gamma}_{+}^{2}\bar{\lambda}_{+}^{2j}+\bar{\gamma}_{-}^{2}\bar{\lambda}_{-}^{2j}\;,
\end{equation}
and \begin{equation}
g_{j}^{m-m'}=\bar{\gamma}_{-}\bar{\gamma}_{+}\bar{\lambda}_{+}^{j}\bar{\lambda}_{-}^{j}\Big(\bar{\lambda}_{-}^{m-m'}+\bar{\lambda}_{+}^{m-m'}\Big).
\end{equation}
After evaluating the summation in above equations (see Appendix~\ref{AppendixB} for detailed derivations) we can approximate 
\begin{equation}
  \braket{p_n^2}_\text{det}=\frac{1}{\bar{\gamma}}\Bigg[\frac{\sqrt{8}\eta\beta\mathcal{C}}{\Omega\tau}\Bigg]^{2}\Bigg[\frac{\sinh^{2}\Big(\frac{\bar{\gamma}_{+}\tau}{4}\Big)}{\bar{\gamma}_{+}(\bar{\gamma}_{+}-\bar{\gamma}_{-})}-\frac{\sinh^{2}\Big(\frac{\bar{\gamma}_{-}\tau}{4}\Big)}{\bar{\gamma}_{-}(\bar{\gamma}_{+}-\bar{\gamma}_{-})}\Bigg]
\end{equation}
This tells us how the effective temperature is modified via the measurement back-action. One needs to then optimize $\tau$ such that cooling rate is still high while the noise is minimized. Further, we will assume here equipartition between the variance of the position and the momentum and then derive the effective occupancy $\bar n=\braket{p_n^2}$, and the corresponding effective center of mass temperature is governed as $\bar{T}=\hbar\Omega/[k_{B}\ln(1+1/\bar{n})]$. Finally, one can list the fully analytical expression of the final occupancy as
\begin{equation}
	\bar{n}=\frac{\gamma n_\text{th}}{\gamma+\bar{\Gamma}}\Big(\frac{\Omega}{\bar{\Omega}}\Big)+\frac{1}{\gamma+\bar{\Gamma}}\Bigg[\frac{\sqrt{8}\eta\beta\mathcal{C}}{\Omega\tau}\Bigg]^{2}\Bigg[\frac{\mathcal{A}_{+}-\mathcal{A}_{-}}{\bar{\gamma}_{+}-\bar{\gamma}_{-}}\Bigg]\;,
	\label{FinalOccupancy}
\end{equation}
where we introduce new notations as
\begin{equation}
\mathcal{A}_{\pm}=\frac{1}{\bar{\gamma}_{\pm}}\sinh^{2}\Big(\frac{\bar{\gamma}_{\pm}\tau}{4}\Big)\;.
\end{equation}
Notice that in the limit of $\tau\rightarrow 0$, the expression above simplifies to
\begin{align}
	\bar{n}\approx\frac{\gamma n_\text{th}}{\gamma+\beta\mathcal{C}}+\frac{\Big(\frac{\eta\beta\mathcal{C}}{\sqrt{2}\Omega}\Big)^{2}}{\gamma+\beta\mathcal{C}}\;.\label{FinalOccupancy_lowtau}
\end{align}
In this limit, $\bar{\Omega}\approx\Omega$ and one recovers a standard result in optomechanics where the thermal occupancy is reduced to the ratio of the heating $\gamma n_\text{th}$ to the externally imposed cooling rate added to the intrinsic damping rate $\gamma+\beta\mathcal{C}$. The second term is the inadvertently added feedback noise, scaling with $\eta^2$. We will discuss the implications of this expression in the following section.

%%%%%%%%%%%%%%%%%%%%%%%%%%%%%%%%%%%%%%%%%%
%%%%%%%%%%%%%%%%%%%%%%%%%%%%%%%%%%%%%%%%%%
\section{Discussions}
\label{sec4}
%%%%%%%%%%%%%%%%%%%%%%%%%%%%%%%%%%%%%%%%%%
%%%%%%%%%%%%%%%%%%%%%%%%%%%%%%%%%%%%%%%%%%
Let us address here the feasibility of such a scheme to cool the classical thermal motion of a magnetic micro-sphere levitated inside the optical hollow-core fiber. To this end we connect to the experimental setup described in Ref.~ \cite{chakraborty2024arXiv}, making use of some of the parameters relevant for current experiments. We optimize the resulting expression for the final occupancy by taking into account the influence of gas pressure and we describe a procedure for feedback implementation. The validity of the analytical estimates is checked against numerical simulations of the stochastic difference equations.\\

%%%%%%%%%%%%%%%%%%%%%%%%%%%%%%%%%%%%%%%%%%%%%%%%%%%%%%%%%%%%%%%%%%%%%%
%%%%%%%%%%%%%%%%%%%%%%%%%%%%%%%%%%%%%%%%%%%%%%%%%%%%%%%%%%%%%%%%%%%%%%
\subsection{Optimizing the final occupancy}
%%%%%%%%%%%%%%%%%%%%%%%%%%%%%%%%%%%%%%%%%%%%%%%%%%%%%%%%%%%%%%%%%%%%%%
%%%%%%%%%%%%%%%%%%%%%%%%%%%%%%%%%%%%%%%%%%%%%%%%%%%%%%%%%%%%%%%%%%%%%%
In a setup described in Appendix \ref{AppendixC}, the magnetic microparticle can be stably trapped until around $1$ mbar. However, after this it escapes from the trap due to an increase in internal bulk temperature of the particle and insufficient damping at low pressure~\cite{ricciThesis2019}. This not only restricts the experiment to be performed at a pressure below $1$ mbar but also limits its quality factor. A possible solution to these limitations can be provided by using the phase-adaptive feedback operation in order to keep the particle stable in the optical trap during the vacuum chamber evacuation procedure. As shown in Eq.~(\ref{FinalOccupancy}), the final phonon occupancy in the system is limited by the product $\eta\beta\mathcal{C}$. Experimentally, the parameter $\beta$ can be controlled by the trap stiffness which in turn can be adjusted by parameters such as the trapping power, and core size of the fiber. For the setup in Ref.~\cite{chakraborty2024arXiv}, $\beta=\kappa_{\mathrm{trap}}/(k_{l} p_\mathrm{zpm} )=(\sqrt{m} \Omega^{3/2} \lambda_{l})/(2\pi\sqrt{\hbar})=1.3\times10^{11}$~Hz. The noise floor for our current detection scheme is $17$~pm/$\sqrt{\mathrm{Hz}}$ [see Appendix \ref{AppendixC}] and from this we can obtain the parameter $\eta$. To do so, we take into account of the spectral contribution due to detection noise (see appendix~\ref{AppendixD} for details) and around $\omega\approx\Omega$, we find $\eta=9.2\times 10^{6}~\sqrt{\mathrm{Hz}}$.   The amplifier gain $\mathcal{C}$ can be set to an optimum value in order to avoid subsequent additive noise while maintaining optimum signal-noise ratio during the stages of detection and feedback. To obtain this factor, we take $\tau\rightarrow 0$ in Eq.~(\ref{EqPhase}). This leads to $\phi=\mathcal{C}p_{det}\approx\mathcal{C}\sqrt{n_\text{th}}$, where in a thermal state the variance of the momentum quadrature is taken to be $n_\text{th}$. The upper bound of the detected phase is $2\pi$. Using this, we can estimate the maximum value of the gain parameter as $\mathcal{C}_{max}=2\pi/\sqrt{n_\text{th}}$. On the other hand, the optimum value of $\mathcal{C}$ can be obtained by minimizing the phonon occupation in Eq. (\ref{FinalOccupancy_lowtau}). For $\tau\rightarrow 0$, this gives following optimum value of the amplifier gain
\begin{align}
	\mathcal{C}_{opt}=-\frac{\gamma}{\beta}+\sqrt{\frac{\gamma^{2}}{\beta^{2}}+\frac{2\Omega^{2}\gamma n_{th}}{\eta^2\beta^2}}\;.
\end{align} 
 For an underdamped oscillator $\gamma\ll\beta$, and above expression attains a simplified form, $\mathcal{C}_{opt}\approx\frac{\Omega}{\eta\beta}\sqrt{2\gamma n_{th}}$. A generalized expression of the optimum gain valid for any $\tau$ is provided in Appendix \ref{AppendixE}.
%%%%%%%%%%%%%%%%%%%%%%%%%%%%%%%%%%%%%%%%%%%%%%%%%%%%%%%%%%%%%%%%%%%%%%
%%%%%%%%%%%%%%%%%%%%%%%%%%%%%%%%%%%%%%%%%%%%%%%%%%%%%%%%%%%%%%%%%%%%%%
\subsection{Numerical simulations}
%%%%%%%%%%%%%%%%%%%%%%%%%%%%%%%%%%%%%%%%%%%%%%%%%%%%%%%%%%%%%%%%%%%%%%
%%%%%%%%%%%%%%%%%%%%%%%%%%%%%%%%%%%%%%%%%%%%%%%%%%%%%%%%%%%%%%%%%%%%%%
Under low-vacuum conditions, the motion of trapped particle is strongly damped due to instantaneous collisions with the background gas and its energy quickly thermalizes with the environment at a rate $\gamma$. The momentum damping rate of the levitated particle can be reduced by decreasing the air pressure. From kinetic theory the damping constant follows the relation \cite{HutchinsAST1995,BykovNatPhys2015}
\begin{align}
	\gamma &=\frac{12\pi R}{m}\frac{\mu_{0}}{1+\mathrm{Kn}\Big(\beta_{1}+\beta_{2}e^{-\beta_{3}/\mathrm{Kn}}\Big)}\;,\label{Damping}
\end{align}
where $R$ is the particle radius, and $m$ is its mass and $\mu_{0}=18.1\times 10^{-6}$ Pa.s is the coefficient of viscosity of air, $\beta_{1}=1.231$, $\beta_{2}=0.469$, and $\beta_{3}=1.178$ at the atmospheric pressure. Further in Eq. (\ref{Damping}), the Knudsen number $\mathrm{Kn}=\frac{\lambda_{\mathrm{mfp}}}{R}\propto\frac{1}{pR}$ is expressed as a ratio  of the mean free path of the background gas to the radius of the particle, where $p$ is the pressure of the surrounding gas. It can be verified from Eq. (\ref{Damping}) that the damping of the trapped particle is independent of pressure in high-pressure regime where its motion gets over-damped due to strong interaction with the background gas. Contrary to this, for low-pressure, the mean free path becomes much larger than the radius of the particle and the damping in this regime varies linearly with pressure.
%%%%%%%%%%%%%%%%%%%%%%%%%%%%%%%%%%%%%%%%%%%%%%%%%%%%%%%%%%%%%%%%%%%%%%
%%%%%%%%%%%%%%%%%%%%%%%%%%%%%%%%%%%%%%%%%%%%%%%%%%%%%%%%%%%%%%%%%%%%%%
\begin{figure}[t]
	\includegraphics[width=0.98\columnwidth]{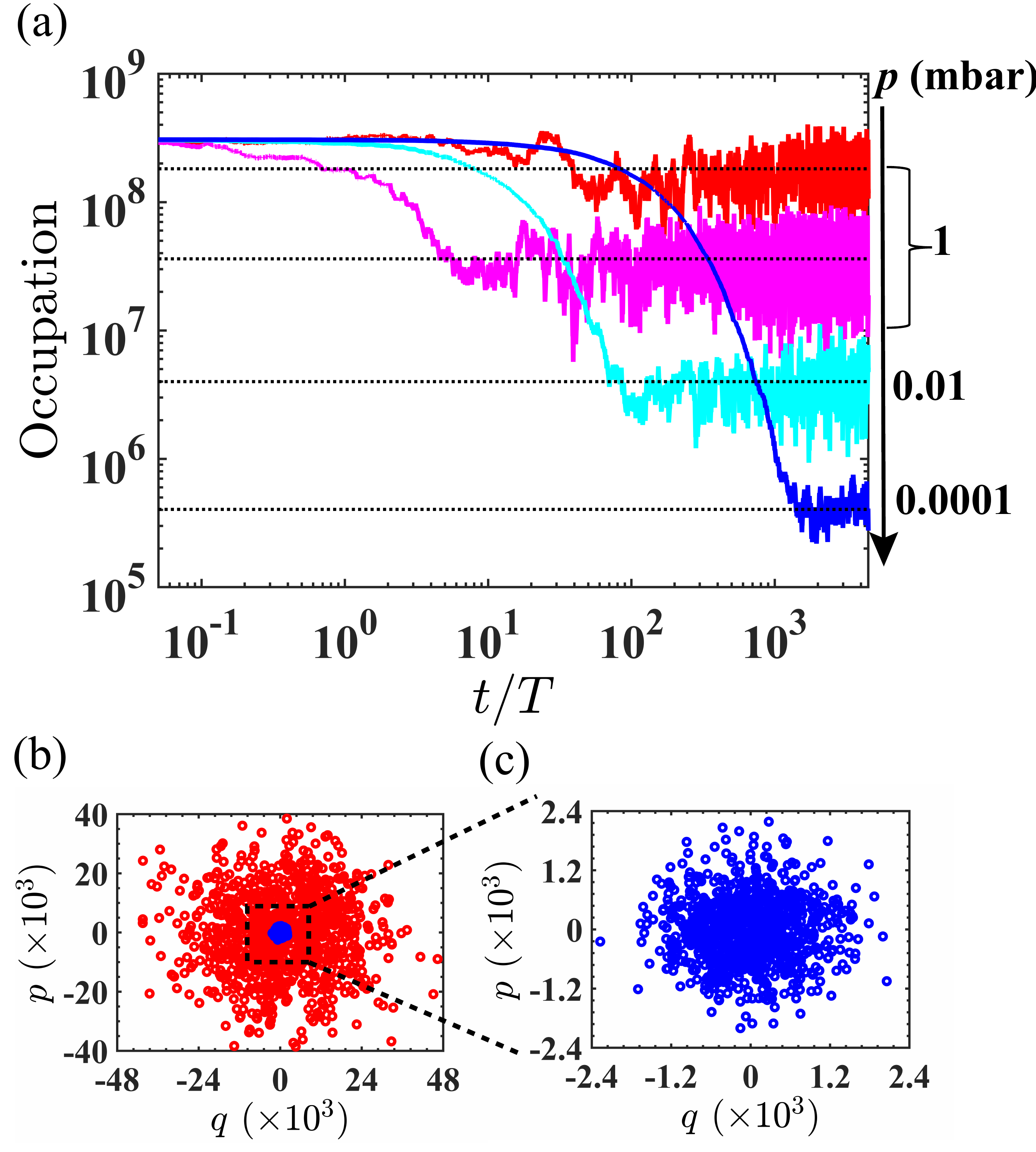}
	\caption {Phase adaptive feedback cooling. (a) Temporal evolution of the phonon occupancy (log-log scale) for different pressure levels. The occupancy is obtained from a numerical simulation of Eqs.~(\ref{EqsMotion}), incorporating both thermal and detection noises, along with the pressure dependent damping behavior governed by Eq.~ (\ref{Damping}). The dotted lines show the validity of the analytically predicted final occupancy given by Eq. (\ref{FinalOccupancy}). In the absence of feedback cooling (solid red line), the system quickly thermalizes with the environment at 1 mbar, leading to high phonon-occupancy. However, with the feedback activation, the phonon occupancy is reduced, as depicted for pressures of 1 mbar (solid magenta line), 10 $\mu$bar (solid cyan line), and 0.1 $\mu$bar (solid blue line).  (b) Phase-space representation of an initial thermal state (red points) and the cooling state (blue points) obtained for $p=0.1~\mu$bar, and (c) for a corresponding pressure, enlarged view of the thermal distribution of the final cold state. Parameters used are   $\Omega/2\pi=34.5$ kHz, $T=2\pi/\Omega$,  $\tau/T=7.9\times 10^{-4}$, $T_{\mathrm{th}}=300$ K, $n_{\mathrm{th}}=1.8 \times 10^{8}$, $\beta/2\pi=20.6$ GHz, $\mathcal{C}=\mathcal{C}_{opt}$, $\eta=9.2\times 10^{6}\sqrt{\mathrm{Hz}}$, $\mu_{0}=18.1$ Pa.s, density of the trapped particle = 5.11 g/$\mathrm{cm}^{3}$, and $R=0.66~\mu$m (see Appendix \ref{AppendixC1} for discussion on Mie scattering).}
	\label{fig4}
\end{figure}
%%%%%%%%%%%%%%%%%%%%%%%%%%%%%%%%%%%%%%%%%%%%%%%%%%%%%%%%%%%%%%%%%%%%%%
%%%%%%%%%%%%%%%%%%%%%%%%%%%%%%%%%%%%%%%%%%%%%%%%%%%%%%%%%%%%%%%%%%%%%%
Let us now refer to full numerical simulations corresponding to small $\tau$. We show in Fig.~\ref{fig4} sample trajectories obtained from integrating the set of Eqs.~ \eqref{EqsMotion} in the presence of both thermal and feedback noise modeled using Wiener increments (see Appendix~\ref{AppendixF} for details of numerical procedure). The phonon occupancy is plotted for $\tau/T=7.9\times 10^{-4}$ and for different pressures. In the absence of external damping and at a pressure $p=1$ mbar, the phonon-occupancy is dominated by the stochastic Brownian thermal noise and the oscillator remains in equilibrium with its thermal environment, as depicted in Fig.~\ref{fig4}(a). When the phase-adaptive feedback cooling is switched on, the initial high phonon occupancy decays at the analytically predicted cooling rate $\gamma+\bar{\Gamma}$ and eventually saturates at a final value $\bar{n}$ governed by and validating our analytical expression in Eq.~(\ref{FinalOccupancy}). The equilibrium is reached due to the competition between the thermal reheating and the measurement backaction. 

As the pressure is reduced further, the initial high thermal phonon occupancy decays further albeit at a slower rate as depicted in Fig.~\ref{fig4}. For instance, at $0.1$ $\mu$bar of pressure, the phonon population can be reduced to the level of $10^{5}$ from the initial high occupation of $10^{8}$. 

In terms of effective temperature, at a pressure of $1$ mbar, our formalism provides a translational motion temperature of approximately 60 K. At a comparable pressure, this is notably higher than the minimal effective temperature of 5.87 K reported in \cite{GrassAPL2016}, which was achieved for a relatively smaller silica particle of size 387 nm. However, pressure reduction in our scheme results in significantly lower temperatures. For instance, at $0.1~\mu$bar, an effective temperature of 0.7 K can be achieved using our proposed feedback technique. With further improved high vacuum technology, our technique has a potential to approach temperatures in the millikelvin range.

The reduction in thermal energy can further be understood in terms of phase-space representation plotted in Figs.~\ref{fig4}(b) and (c) for a pressure of 0.1 $\mu$bar.  Due to the applied feedback, the thermal distribution of variance $n_{\mathrm{th}}$ corresponding to an initial hot state suppresses in phase-space to an analogous distribution of reduced variance $\bar{n}$ associated with a final cold state. Notice that the phase space result validate the assumption that equipartition applies even in the presence of externally controlled damping.
%%%%%%%%%%%%%%%%%%%%%%%%%%%%%%%%%%%%%%%%%%%%%%%%%%%%%%%%%%%%%%%%%%%%%%
%%%%%%%%%%%%%%%%%%%%%%%%%%%%%%%%%%%%%%%%%%%%%%%%%%%%%%%%%%%%%%%%%%%%%%
\begin{figure}[t]
	\includegraphics[width=0.8\columnwidth]{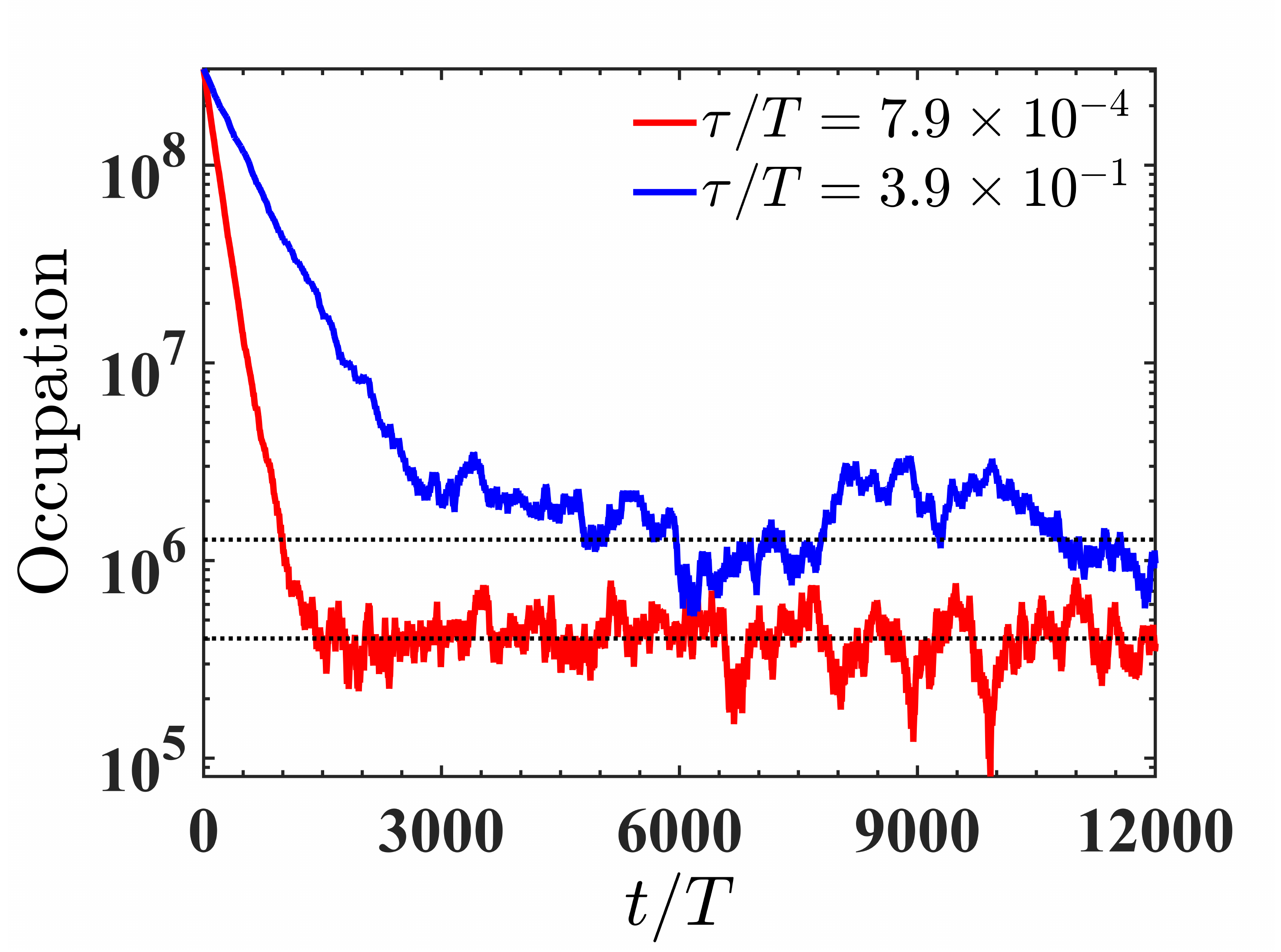}
	\caption {Plot of the phonon occupancy as a function of time at $0.1$~$\mu$bar for $\tau/T=7.9\times 10^{-4}$ (solid magenta line) and $\tau/T=3.9\times 10^{-1}$ (solid cyan line). The results are extracted from numerical simulations of Eqs.~(\ref{EqsMotion}). The dotted lines illustrate the analytically predicted final occupancy values. Other parameters used are same as in Fig.~\ref{fig4}.}
	\label{fig5}
\end{figure}
%%%%%%%%%%%%%%%%%%%%%%%%%%%%%%%%%%%%%%%%%%%%%%%%%%%%%%%%%%%%%%%%%%%%%%
%%%%%%%%%%%%%%%%%%%%%%%%%%%%%%%%%%%%%%%%%%%%%%%%%%%%%%%%%%%%%%%%%%%%%%

For larger values of $\tau$, the scheme describes a quasi-derivative feedback regime instead of an instantaneous cold-damping operation. At $0.1$~$\mu$bar, in Fig.~\ref{fig5}, we compare the phonon occupation obtained from two feedback operations for $\tau/T=7.9\times 10^{-4}$ and $\tau/T=3.9\times 10^{-1}$. Interestingly, the quasi-derivative scheme performs well albeit the final phonon occupation remains slightly higher than the cold damping. It is due to the increase in detection noise over longer time-intervals of feedback operation. Further, for delayed feedback, the thermal energy decays at a slower rate than cold-damping scheme. These remarks suggest that in our phase-adaptive mechanism, the effective Stokes-type damping force (when $\tau\rightarrow 0$) offers a best strategy for the isolation from  thermal bath.\\

In the preceding analysis, we illustrated axial (one-dimensional) feedback cooling via phase control to mitigate the dominant instability observed in experiments \cite{chakraborty2024arXiv}. While the fiber mode’s intensity profile offers intrinsic transverse confinement, full trap stability requires active transverse stabilization. Parametric methods like transverse phase modulation \cite{ghosh2023theory} or intensity modulation \cite{gieseler2012subkelvin} could extend stabilization to all three dimensions. A synergistic approach combining cold damping via phase control with  phase-adaptive parametric method \cite{ghosh2023theory}  shows potential for robust axial as well as transverse stabilization by harnessing the complementary strengths of both techniques. Furthermore, the incorporation of electrical cooling techniques \cite{conanglaPRL2019,goldwaterQST2019} in the radial direction may further improve system performance and robustness.

\subsection{Internal temperature of the levitated particle}
 So far, we have described the cooling strategy to effectively reduce the thermal phonon occupation. In other words, this leads to a reduction in the center of mass temperature of the levitated particle. However, at low pressures where the background gas density is minimal, levitated particles have issues with stability, owing to their internal heating \cite{MillenNatNanoTech2014}.  For submicron particles, the steady state internal temperature at low pressure is primarily determined by the following competing processes: (i) heating due to optical absorption of trapping light and blackbody radiation, and (ii) cooling through the emission of blackbody radiation. These factors can lead to extremely high bulk temperatures. For instance, for silica of radius 68 nm levitated in free space, the internal temperature can often exceed 1000 K \cite{HebestreitPRA2018}. However, the polarizability of the trapped magnetic micron-sized particle in our system is unfortunately not available at the peak of the blackbody radiation, making it difficult to precisely estimate its internal temperature. Nonetheless, assuming that the polarizability is in the same ballpark as for silica, for a magnetic trapped particle of radius of around $660$~nm levitated with optical intensity 2.4 $\times~10^{11}$ W/m$^2$, an  estimate of the internal temperature suggest 642 K times a factor reflecting the difference between the polarizabilities.
 
 Interestingly, the issue of internal temperature can be tackled by a mechanism to redistribute energy via spontaneous emission at selected resonances of large optical emission rate. This is an Anti-stokes cooling technique used in the field of cooling, or rather said \textit{refrigeration } of solid-state objects \cite{RahmanNatPhoton2017}.  The Anti-stokes cooling can be implemented using particles doped with rear earth elements that show very sharp optical resonances \cite{BensalahOptMat2004}. The thermal emission of such particles into free space allows efficient redistribution of the thermal energy leading to a possible reduction in the internal temperature  alongwith ensuring the stability of the particle in optical trap and allowing operation under extremely low pressure.

\subsection{Laser frequency noise effects}
\begin{figure}[t]
		\includegraphics[width=0.8\columnwidth]{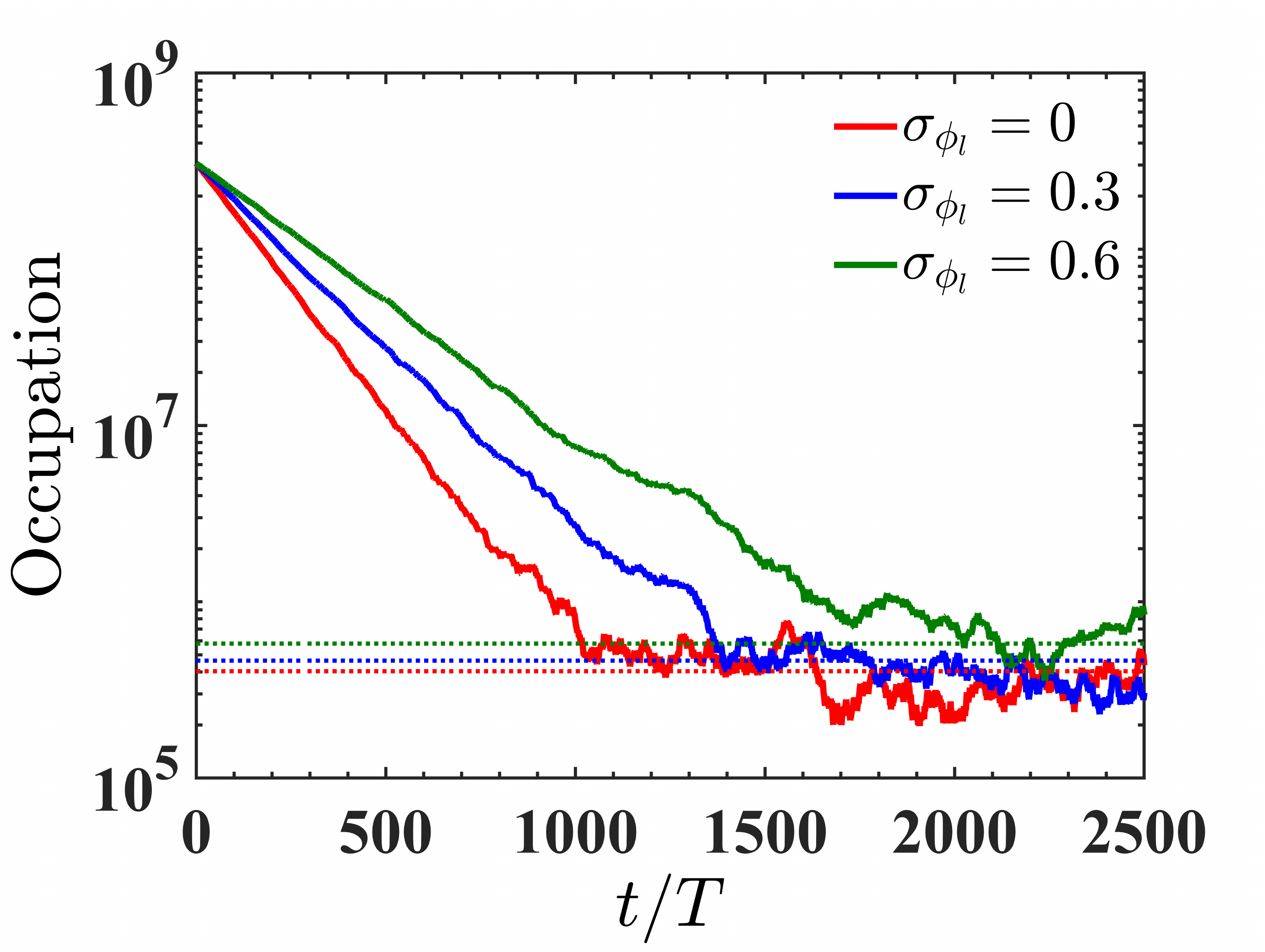}
	\caption {Effect of trapping laser phase noise on system dynamics for a pressure of $0.1~\mu$bar. Solid lines represent simulated trajectories for phase noise variances $\sigma_{\phi_{l}}=0$ (red), 0.3 (blue) and 0.6 (green), corresponding to trap frequency changes of  $0\%$, $\sim 14\%$ and $\sim 26\%$, respectively.  Dotted lines denote the analytically predicted steady-state average occupancy values. All other parameters are same as in Fig.~\ref{fig4}.}
		\label{fig6}
	\end{figure}

Let us now consider the effect of fluctuations in the trapping laser frequency, which arise due to uncertainty in the laser phase. We model the noisy laser field as having a fixed amplitude and a time-dependent fluctuating phase \cite{rablPRA2009}, so that  $\mathcal{E}_{0}=|\mathcal{E}_{0}|e^{i\phi_{l}(t)}$, where $\phi_{l}$ is a random Gaussian variable of zero mean and variance $\sigma_{\phi_{l}}$. The effect of this phase noise (see Appendix \ref{AppendixG} for details) can be captured by averaging the total electric field, $\langle \mathcal{E}(y,t)\rangle$. Using the property of a Gaussian noise, we have $\langle e^{i\phi_{l}(t)}\rangle= e^{-\sigma_{\phi_{l}}/2}$, where averaging is taken over many realizations. By applying error propagation techniques to the governing equations, we find that the phase uncertainty rescales the trapping frequency from $\Omega$ to $\Omega e^{-\sigma_{\phi_{l}}/2}$ and reduces the effective cooling rate from $\bar{\Gamma}$ to $\bar{\Gamma}e^{-\sigma_{\phi_{l}}}$. In this approach, the detection noise amplitude $\eta$ and optimized gain factor $\mathcal{C}_{opt}$ are considered independent from the phase noise contributions. The impact of laser phase noise is shown in Fig. \ref{fig6} for different values of $\sigma_{\phi_{l}}$. The ensemble-averaged trajectories align precisely with the analytical predictions, demonstrating a systematic decrease in cooling rate as phase noise variance grows. This relationship is directly reflected in the negative slope of the curves shown in Fig. \ref{fig6}. Beyond its role in reducing the cooling rate, phase noise has only a marginal impact—even when combined with thermal and detection noise—resulting in an almost negligible change in the final phonon occupancy.
	%%%%%%%%%%%%%%%%%%%%%%%%%%%%%%%%%%%%%%%%%%%%%%%%%%%%%%%%%%%%%%%%%%%%%%
	%%%%%%%%%%%%%%%%%%%%%%%%%%%%%%%%%%%%%%%%%%%%%%%%%%%%%%%%%%%%%%%%%%%%%%
	%%%%%%%%%%%%%%%%%%%%%%%%%%%%%%%%%%%%%%%%%%%%%%%%%%%%%%%%%%%%%%%%%%%%%%
	%%%%%%%%%%%%%%%%%%%%%%%%%%%%%%%%%%%%%%%%%%%%%%%%%%%%%%%%%%%%%%%%%%%%%%

%%%%%%%%%%%%%%%%%%%%%%%%%%%%%%%%%%%%%%%%%%%%%%%%%%%%%%%%%%%%%%%%%%%%%%%
%%%%%%%%%%%%%%%%%%%%%%%%%%%%%%%%%%%%%%%%%%%%%%%%%%%%%%%%%%%%%%%%%%%%%%%
\section{Conclusions}
\label{sec5}
%%%%%%%%%%%%%%%%%%%%%%%%%%%%%%%%%%%%%%%%%%%%%%%%%%%%%%%%%%%%%%%%%%%%%%
%%%%%%%%%%%%%%%%%%%%%%%%%%%%%%%%%%%%%%%%%%%%%%%%%%%%%%%%%%%%%%%%%%%%%%
We presented the prospects of cooling the thermal excess fluctuations linked to the mechanical motion of a magnetic microparticle, optically levitated inside a hollow-core photonic crystal fiber and in the presence of a thermal gas. The cooling is achieved through an engineered phase-adaptive feedback loop that provides an effective Stokes type viscous damping force to the system. This is accomplished by directly imaging the particle's position followed by the adjustment of relative phase between two counter-propagating fiber guided waves. We provided an analytical estimate of the final phonon occupation taking into account of thermal as well as the measurement noises. Our numerical simulations validate the analytical results. Our findings are consequential for using trapped magnetic particles as sensors for feeble stray magnetic fields~\cite{ZareRameshtiPhysRep2022}. Future work, analysing the feasibility of reaching the quantum ground state, will aim at extending this procedure to prepare the non-classical states of magnetization \cite{sharmaPRB2021, wachterJOSAB2021, magriniPRL2022} and testing aspect of fundamental quantum physics \cite{chigusaPRD2020, arndtNatPhys2014}.

%%%%%%%%%%%%%%%%%%%%%%%%%%%%%%%%%%%%%%%%%%%%%%%%%%%%%%%%%%%%%%%%%%%%%%
%%%%%%%%%%%%%%%%%%%%%%%%%%%%%%%%%%%%%%%%%%%%%%%%%%%%%%%%%%%%%%%%%%%%%%
 \section*{Acknowledgments}
 We acknowledge financial support from the Max Planck Society and the Deutsche Forschungsgemeinschaft (DFG, German Research Foundation) -- Project-ID 429529648 -- TRR 306 QuCoLiMa
(``Quantum Cooperativity of Light and Matter'').
%\bibliography{Refs}


\appendix
\begin{widetext}
%%%%%%%%%%%%%%%%%%%%%%%%%%%%%%%%%%%%%%%%%%%%%%%%%%%%%%%%%%%%%%%%%%%%%%
%%%%%%%%%%%%%%%%%%%%%%%%%%%%%%%%%%%%%%%%%%%%%%%%%%%%%%%%%%%%%%%%%%%%%%
%%%%%%%%%%%%%%%%%%%%%%%%%%%%%%%%%%%%%%%%%%%%%%%%%%%%%%%%%%%%%%%%%%%%%%
%%%%%%%%%%%%%%%%%%%%%%%%%%%%%%%%%%%%%%%%%%%%%%%%%%%%%%%%%%%%%%%%%%%%%%
\section{Analytical solution for cooling rate}
\label{AppendixA}
We start with the deterministic part of Eqs.~\eqref{EqsMotion} while neglecting the effect of noise
\begin{subequations}
\begin{align}
\frac{dq}{dt} &= \Omega p\;\\
\frac{dp}{dt} &=-\gamma p-\Omega q-\beta \phi(t).
\end{align}\label{EqsMotionNoNoise_AppendixA}
\end{subequations}
Formally integrating the first equation gives
\begin{equation}
q(t)-q(t-\tau)=\Omega\int_{t-\tau}^{t}p(t')dt'.
\end{equation}

Let us assume that the feedback is not affecting strongly over the time interval $t-\tau$ to $t$. Then, to a first order, one can take into account that the natural evolution occurs at frequency $\Omega$ and damping rate $\gamma$. This allows one to deterministically connect $p(t')$ with $p(t)$ and $q(t)$

\begin{align}
	p(t')= \frac{\Omega}{\tilde{\Omega}}\Big[e^{(\gamma+\tilde{\Omega})(t-t')/2}-e^{(\gamma-\tilde{\Omega})(t-t')/2}\Big]p(t)+\frac{1}{2\tilde{\Omega}}\Big[(\gamma+\tilde{\Omega})e^{(\gamma+\tilde{\Omega})(t-t')/2}-(\gamma-\tilde{\Omega})e^{(\gamma-\tilde{\Omega})(t-t')/2}\Big]q(t)\;,
\end{align}
where $\tilde{\Omega}=\sqrt{\gamma^{2}-4\Omega^{2}}$. After performing the integration of decaying, oscillating terms, one sees a renormalization of the decay rate from $\gamma$ to $\gamma+\bar{\Gamma}$ (by identifying the terms proportional to $p(t)$) where

\begin{equation}
	\bar{\Gamma} =\beta \mathcal{C}\Bigg[e^{\gamma\tau/2}~\frac{\sinh{\tilde{\Omega} \tau/2}}{\tilde{\Omega}\tau/2}\Bigg]\;.
\end{equation}

In addition, the natural frequency is also modified from $\Omega$ to $\bar{\Omega}$ and can be identified from the extra factors in front of $q(t)$ leading to

\begin{equation}
	\bar \Omega =\Omega-\beta \mathcal{C}\left[\frac{e^{\gamma \tau/2}\cosh{(\tilde{\Omega} \tau/2)}-1}{\Omega \tau}\right]+\frac{\gamma\bar{\Gamma}}{2\Omega}\;.
\end{equation}
In the limit of $\tau\rightarrow 0$ i.e. when the derivative of the momentum is used to modify the motion, we have expected cold-damping type of result $\bar{\Gamma}=\Gamma=\beta\mathcal{C}$ and the corresponding frequency of oscillation $\bar \Omega=\Omega$. \\

For an under-damped oscillator $\gamma\ll\Omega$, the expressions for the modified optical damping and frequency govern the following form
\begin{align}
		\bar{\Gamma}&=\beta \mathcal{C}\Bigg[e^{\gamma\tau/2}~\frac{\sin{\Omega \tau}}{\Omega\tau}\Bigg]\;,\\
		\bar \Omega&=\Omega-\beta \mathcal{C}\left[\frac{e^{\gamma \tau/2}\cos(\Omega \tau)-1}{\Omega \tau}\right]\;.
\end{align}

%%%%%%%%%%%%%%%%%%%%%%%
\begin{figure}[t]
	\includegraphics[width=0.5\columnwidth]{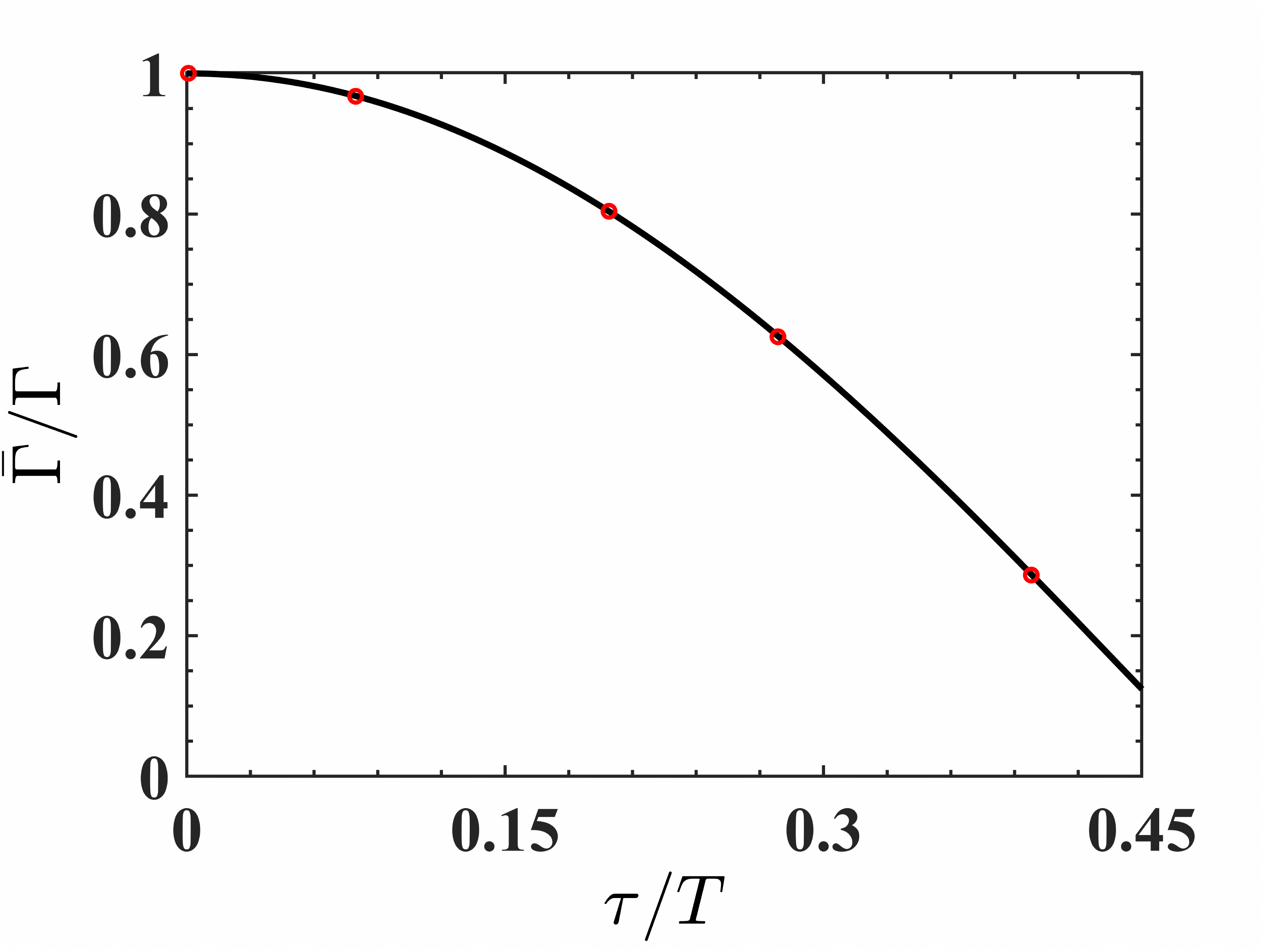}
	\caption {Variation of the modified  optical damping rate  as a function of $\tau/T$ where $T=2\pi/\Omega$ and $\Omega=2\pi\times 34.5$ kHz. The parameters used are $p=1$ mbar, $\beta/2\pi=20.6$ GHz, $\mathcal{C}=\mathcal{C}_{opt}^{\tau}$. The analytical result and a few numerical checks shown as red dots confirm the predicted behavior.}
	\label{figAppendixA}
\end{figure}
%%%%%%%%%%%%%%%%%%%%%%%
In Fig. \ref{figAppendixA}, we depict the modified optical damping as a function of $\tau/T$. The predicted behavior of the modified damping rate can be checked by deducing its values for different $\tau$ from the occupancy obtained with the numerical solution of Eqs. (\ref{EqsMotionNoNoise_AppendixA}). This is confirmed by the red circles shown in Fig. \ref{figAppendixA}. 

%%%%%%%%%%%%%%%%%%%%%%%%%%%%%%%%%%%%%%%%%%%%%%%%%%%%%%%%%%%%%
%%%%%%%%%%%%%%%%%%%%%%%%%%%%%%%%%%%%%%%%%%%%%%%%%%%%%%%%%%%%%
\section{Classical stochastic evolution: thermal and measurement noise}
\label{AppendixB}
%%%%%%%%%%%%%%%%%%%%%%%%%%%%%%%%%%%%%%%%%%%%%%%%%%%%%%%%%%%%%%%%%%%%%%%%%%%%%%%%%%%%%%%%%%
%%%%%%%%%%%%%%%%%%%%%%%%%%%%%%%%%%%%%%%%%%%%%%%%%%%%%%%%%%%%%%%%%%%%%%%%%%%%%%%%%%%%%%%%%%
We first analyze the thermal noise in order to get the hang of it. Afterwards, we will show how to compute the final occupancy by correlations of the measurement noise.
\subsection{Thermal noise}
The coupled difference equations for a harmonic oscillator undergoing damping
\begin{subequations}
  \begin{align}
  \label{Wiener}
  dq &= \Omega p dt\;,\\
  dp &=-\gamma p dt- \Omega q dt+\sqrt{2\gamma n_\text{th}}dW(t)\;,
  \end{align}
\end{subequations}
can be turned into a set of recurrence equations. We discretize the time interval $[0,t]$ into $n$ steps of duration $dt = t/n$ and can then rewrite the equations above as
\begin{equation}
\mathbf{v}_n - \mathcal{M}\mathbf{v}_{n-1} =\sqrt{2\gamma n_\text{th}} \mathbf{u} dW_{n-1}\;,
\end{equation}
where $\mathbf{v}=(q,p)^\top$ and $\mathbf{u}=(0,1)^\top$ and the evolution matrix is defined as
\begin{equation}
\mathcal{M}=\begin{bmatrix}
1 & \Omega dt \\
-\Omega dt & 1-\gamma dt
\end{bmatrix}\;.
\end{equation}
The eigenvalues and eigenvectors of $\mathcal{M}$ are
\begin{align}
	\lambda_{\pm}&=1-\frac{\gamma_{\pm} dt}{2};~~\alpha_{\pm}=\begin{bmatrix}
		-\frac{\gamma_{\pm} }{2\Omega }\\1\end{bmatrix}\;,
\end{align}
where $\gamma_{\pm}=\gamma\Bigg[1\pm\sqrt{1-\frac{4\Omega^{2}}{\gamma^2}}\Bigg]$. Now the diagonalization of the matrix $\mathcal{M}=S\Lambda S^{-1}$ is straightforward in terms of its eigenvalues.
The resulting quadratures after $n$ steps are written as
\begin{subequations}
  \begin{align}
  \label{qn-pn}
 q_{n}&=\Bigg(\frac{\gamma_{+}\lambda_{-}^{n}-\gamma_{-}\lambda_{+}^{n}}{\gamma_{+}-\gamma_{-}}\Bigg)q_{0}+\Bigg(\frac{\gamma_{-}\gamma_{+}(\lambda_{-}^{n}-\lambda_{+}^{n})}{2\Omega(\gamma_{+}-\gamma_{-})}\Bigg)p_{0}+\Bigg(\frac{\gamma_{-}\gamma_{+}}{2\Omega(\gamma_{+}-\gamma_{-})}\Bigg)\sqrt{2\gamma n_\text{th}}\sum_{j=0}^{n-1}(\lambda_{-}^{j}-\lambda_{+}^{j})dW_{j}\;\\
 p_{n}&=\Bigg(\frac{2\Omega\Big(\lambda_{+}^{n}-\lambda_{-}^{n}\Big)}{\gamma_{+}-\gamma_{-}}\Bigg)q_{0}+\Bigg(\frac{ \gamma_{+}\lambda_{+}^{n}-\gamma_{-}\lambda_{-}^{n}}{\gamma_{+}-\gamma_{-}}\Bigg)p_{0}+\Bigg(\frac{1}{\gamma_{+}-\gamma_{-}}\Bigg)\sqrt{2\gamma n_\text{th}}\sum_{j=0}^{n-1}( \gamma_{+}\lambda_{+}^{j}-\gamma_{-}\lambda_{-}^{j})dW_{j}\;,
\end{align}
\end{subequations}
where the deterministic parts describe simply the oscillatory weakly damped transient evolution and the last terms are the effect of the thermal environment. In the large $n$ limit we find a closed expression
\begin{subequations}
  \begin{align}
  \label{q-p}
  q(t)&=\Bigg(\frac{\gamma_{+}e^{-\gamma_{-}t/2}-\gamma_{-}e^{-\gamma_{+}t/2}}{\gamma_{+}-\gamma_{-}}\Bigg)q_{0}+\Bigg(\frac{\gamma_{-}\gamma_{+}(e^{-\gamma_{-}t/2}-e^{-\gamma_{+}t/2})}{2\Omega(\gamma_{+}-\gamma_{-})}\Bigg)p_{0}+\Bigg(\frac{\gamma_{-}\gamma_{+}}{2\Omega(\gamma_{+}-\gamma_{-})}\Bigg)\sqrt{2\gamma n_\text{th}}\sum_{j=0}^{n-1}(\lambda_{-}^{j}-\lambda_{+}^{j})dW_{j}\;,\\
  p(t)&=\Bigg(\frac{2\Omega\Big(e^{-\gamma_{+}t/2}-e^{-\gamma_{-}t/2}\Big)}{\gamma_{+}-\gamma_{-}}\Bigg)q_{0}+\Bigg(\frac{ \gamma_{+}e^{-\gamma_{+}t/2}-\gamma_{-}e^{-\gamma_{-}t/2}}{\gamma_{+}-\gamma_{-}}\Bigg)p_{0}+\Bigg(\frac{1}{\gamma_{+}-\gamma_{-}}\Bigg)\sqrt{2\gamma n_\text{th}}\sum_{j=0}^{n-1}( \gamma_{+}\lambda_{+}^{j}-\gamma_{-}\lambda_{-}^{j})dW_{j}\;,
  \end{align}
\end{subequations}
where we have used that $\lim_{n\to\infty}(1-\gamma_{\pm} t/2n)^n=e^{-\gamma_{\pm} t/2}$. From these expressions, one can estimate that in steady state $\braket{q^2}_\text{ss}=\braket{p^2}_\text{ss}=n_\text{th}$ by using $\braket{dW_j dW_{j'}}=\delta_{jj'}t/n$ and evaluating the limit
\begin{subequations}
	\begin{align}
  	\lim_{n\rightarrow\infty}\sum_{j=0}^{n-1}\lambda_{\pm}^{2j}&=\frac{1}{\gamma_{\pm}dt}\;,\\
  	\lim_{n\rightarrow\infty}\sum_{j=0}^{n-1}\lambda_{+}^{j}\lambda_{-}^{j}&=\frac{2}{(\gamma_{+}+\gamma_{-})dt}\;.
  \end{align}
\end{subequations}

\subsection{Measurement noise}
To this end we start with the noise propagating via $-\beta \phi(t) dt$ in Eqs.~\eqref{EqsMotion}. This measurement backaction can be written as a Wiener process in the difference equations form as
\begin{equation}
\mathcal{N}^\text{det}(t) = -\frac{\eta \beta \mathcal{C}}{\Omega} \left[\frac{W^{\text{det}}(t)-W^{\text{det}}(t-\tau)}{\tau}\right]dt\;.
\end{equation}
and to be added to the standard thermal noise $\sqrt{2\gamma n_{\mathrm{th}}}dW^{\text{th}}(t)$. The above Wiener process can further be expressed as a discrete sum over Wiener increments with infinitesimal increment of time step $dt$ such that $t=ndt$ and $\tau=n_\tau dt$, we can express the measurement noise as
\begin{equation}
\mathcal{N}^\text{det}(t) = -\mathcal{A}\sum_{m=1}^{n_\tau}dW^{\text{det}}_{n-n_\tau+m}\;,
\label{DetectionNoise}
\end{equation}
where we make the notation $\mathcal{A}=\eta \beta \mathcal{C}dt/(\Omega \tau)$. From here, we follow the derivation given in Ref.~\cite{ghosh2023theory} to determine the contribution of thermal and detection noises to the steady solution that can be obtained, for $p_{ss}(t)$, as a limit of $p_n$ for large $n$
\begin{equation}
	p_n=\frac{1}{\bar{\gamma}_{+}-\bar{\gamma}_{-}}\sum_{j=0}^{n-1}\Big(\mathcal{N}_{j}^{det}+\sqrt{2\gamma n_\text{th}}dW^{th}_{j}\Big)\Big(\bar{\gamma}_{+}\bar{\lambda}_{+}^{j}-\bar{\gamma}_{-}\bar{\lambda}_{+}^{j}\Big)\;,
\end{equation}
where  $\bar{\lambda}_{\pm}=1-\bar{\gamma}_{\pm}dt/2$ and $\bar{\gamma}_{\pm}=\bar{\gamma}\Big[1\pm\sqrt{1-\frac{4\bar{\Omega}^{2}}{\bar{\gamma}^2}}\Big]$ contain a modified damping $\bar{\gamma}=\gamma+\bar{\Gamma}$, and frequency $\bar{\Omega}$.  Now our task is to compute the variance $\braket{p_n^2}$ in steady state. For two uncorrelated noises one can split this into $\braket{p_n^2}=\braket{p_n^2}^{\text{th}}+\braket{p_n^2}^{\text{det}}$.
Let's first estimate the contribution from  thermal noise, $\braket{p_n^2}^{\text{th}}$, as an exercise. We use $\braket{dW_j^{\text{th}}dW_{j'}^{\text{th}}}=dt\delta_{jj'}$ and evaluate sum $\lim_{n\rightarrow\infty}\sum_{j=0}^{n-1} \bar{\lambda}_{\pm}^{2j}=1/(\bar{\gamma}_{\pm} dt)$, and $\lim_{n\rightarrow\infty}\sum_{j=0}^{n-1}\bar{\lambda}_{+}^{j}\bar{\lambda}_{-}^{j}=\frac{2}{(\bar{\gamma}_{+}+\bar{\gamma}_{-})dt}$ (for details see previous section). This gives the expected estimate of the thermal final occupancy
\begin{equation}
\braket{p_n^2}^{\text{th}}=\frac{\gamma n_\text{th}}{\gamma+\bar{\Gamma}}\Big(\frac{\Omega}{\bar{\Omega}}\Big)\;.
\end{equation}
Now to evaluate the contribution of measurement backaction, we write the steady state correlations stemming from the detection noise as
\begin{equation}
\braket{p_n^2}^\text{det}=\frac{1}{(\bar{\gamma}_{+}-\gamma_{-})^2}\sum_{j=0}^{n-1}\sum_{j'=0}^{n-1}\braket{\mathcal{N}^{\mathrm{det}}_j\mathcal{N}^{\mathrm{det}}_{j'}}(\bar{\gamma}_{+}\bar{\lambda}_{+}^{j}-\bar{\gamma}_{-}\bar{\lambda}_{+}^{j})(\bar{\gamma}_{+}\bar{\lambda}_{+}^{j'}-\bar{\gamma}_{-}\bar{\lambda}_{+}^{j'})\;.
\end{equation}
Using Eq. (\ref{DetectionNoise}), this leads to
\begin{equation}
 	\langle p_{n}^{2}\rangle^{\mathrm{det}}=\frac{\mathcal{A}^{2}}{(\bar{\gamma}_{+}-\bar{\gamma}_{-})^{2}}\sum_{j,j^{'}=0}^{n-1}\sum_{m,m^{'}=0}^{n_{\tau}}\Big(\bar{\gamma}_{+}\bar{\lambda}_{+}^{j}-\bar{\gamma}_{-}\bar{\lambda}_{-}^{j}\Big)\Big(\bar{\gamma}_{+}\bar{\lambda}_{+}^{j^{'}}-\bar{\gamma}_{-}\bar{\lambda}_{-}^{j^{'}}\Big)\langle dW_{j-n_{\tau}+m}^{det}dW_{j^{'}-n_{\tau}+m^{'}}^{det}\rangle\;.
\end{equation}
The correlations of the detection noise will impose the constraint $j+m=j'+m'$ and the expression above reduces to
\begin{equation}
  	\langle p_{n}^{2}\rangle^{\mathrm{det}}=\frac{\mathcal{A}^{2}dt}{(\bar{\gamma}_{+}-\bar{\gamma}_{-})^{2}}\sum_{j=0}^{n-1}\sum_{m,m'=0}^{n_{\tau}}\Bigg[\bar{\gamma}_{+}^{2}\bar{\lambda}_{+}^{2j}+\bar{\gamma}_{-}^{2}\bar{\lambda}_{-}^{2j}-\bar{\gamma}_{-}\bar{\gamma}_{+}\bar{\lambda}_{+}^{j}\bar{\lambda}_{-}^{j}\Big(\bar{\lambda}_{-}^{m-m'}+\bar{\lambda}_{+}^{m-m'}\Big)\Bigg]\;.
\end{equation}
Evaluating the summation over $j$ leads to
\begin{equation}
	\langle p_{n}^{2}\rangle^{\mathrm{det}}=\frac{\mathcal{A}^{2}}{(\bar{\gamma}_{+}-\bar{\gamma}_{-})(\bar{\gamma}_{+}+\bar{\gamma}_{-})}\sum_{m,m'=0}^{n_{\tau}}\Bigg[\bar{\gamma}_{+}\bar{\lambda}_{+}^{m-m'}-\bar{\gamma}_{-}\bar{\lambda}_{-}^{m-m'}\Bigg]\;.
\end{equation}
The summation over $m$ and $m'$ gives $\sum_{m,m'=1}^{n_{\tau}}\bar{\lambda}_{\pm}^{m-m'}=\frac{4\bar{\lambda}_{\pm}}{(\bar{\lambda}_{\pm}-1)^{2}}\sinh^{2}(\bar{\gamma}_{\pm}\tau/4)$. Using this and inserting the expressions of $\bar{\lambda}_{\pm}$, $\bar{\gamma}_{\pm}$ and $\mathcal{A}$ and after some algebric simplifications, we obtain following expression for the variance due to the detection noise
\begin{equation}
	\langle p_{n}^{2}\rangle^{\mathrm{det}}=\Bigg[\frac{\eta\beta\mathcal{C}}{2\Omega}\Bigg]^{2}\frac{1}{\bar{\gamma}\sqrt{\bar{\gamma}^{2}-4\bar{\Omega}^{2}}}\Bigg[\bar{\gamma}_{+}\Bigg(\frac{\sinh\Big(\frac{\bar{\gamma}_{+}\tau}{4}\Big)}{\bar{\gamma}_{+}\tau/4}\Bigg)^{2}-\bar{\gamma}_{-}\Bigg(\frac{\sinh\Big(\frac{\bar{\gamma}_{-}\tau}{4}\Big)}{\bar{\gamma}_{-}\tau/4}\Bigg)^{2}\Bigg]\;.\label{Var_Detection}
\end{equation}
Eventually, the total variance can be written
\begin{equation}
	\label{EqFinalOccp}
	\braket{p_n^2}=\frac{\gamma n_\text{th}}{\gamma+\bar{\Gamma}}\Big(\frac{\Omega}{\bar{\Omega}}\Big)+\Bigg[\frac{\eta\beta\mathcal{C}}{2\Omega}\Bigg]^{2}\frac{1}{\bar{\gamma}\sqrt{\bar{\gamma}^{2}-4\bar{\Omega}^{2}}}\Bigg[\bar{\gamma}_{+}\Bigg(\frac{\sinh\Big(\frac{\bar{\gamma}_{+}\tau}{4}\Big)}{\bar{\gamma}_{+}\tau/4}\Bigg)^{2}-\bar{\gamma}_{-}\Bigg(\frac{\sinh\Big(\frac{\bar{\gamma}_{-}\tau}{4}\Big)}{\bar{\gamma}_{-}\tau/4}\Bigg)^{2}\Bigg]\;.
\end{equation}
The second term basically tells us how much the effective temperature is modified via measurement backaction.
\subsubsection{Under-damped oscillator}
Let us obtain a simplified expression of the variance of the measurement back action for the case of under-damped oscillator i.e. $\bar{\gamma}\ll\bar{\Omega}$. In this case, $\bar{\gamma}_{+}=2i\bar{\Omega}$, and $\bar{\gamma}_{-}=-2i\bar{\Omega}$. Using this in Eq. (\ref{Var_Detection}) gives
\begin{equation}
	\langle p_{n}^{2}\rangle^{\mathrm{det}}=\frac{1}{\gamma+\bar{\Gamma}}\times\Bigg[\frac{\eta\beta\mathcal{C}}{\sqrt{2}\Omega}\times\frac{\sin\Big(\bar{\Omega}\tau/2\Big)}{\bar{\Omega}\tau/2}\Bigg]^{2}\;.
\end{equation}
\subsubsection{Over-damped oscillator}
For an over-damped situation, $\bar{\gamma}\gg\bar{\Omega}$. In this case, $\bar{\gamma}_{+}=2\bar{\gamma}$, and $\bar{\gamma}_{-}\approx 0$. The expression for the variance of the detection noise is then governed by
\begin{equation}
	\langle p_{n}^{2}\rangle^{\mathrm{det}}=\frac{1}{\gamma+\bar{\Gamma}}\times\Bigg[\frac{\eta\beta\mathcal{C}}{\sqrt{2}\Omega}\times\frac{\sinh\Big(\bar{\gamma}\tau/2\Big)}{\bar{\gamma}\tau/2}\Bigg]^{2}\;.
\end{equation}

	\section{Experimental considerations}
	\label{AppendixC}
	Let us refer to the setup in Ref.~\cite{chakraborty2024arXiv} where a magnetic microparticle is optically levitated inside a chiral single-ring HC-PCF at low pressure. A schematic of the setup is shown in Fig.~\ref{Fig1AppendixC} along with the insets showing the scanning electron micrographs of the magnetic microparticle, the HC-PCF and snapshot of an optically trapped magnetic particle captured with a high-speed camera.
	
	A linearly polarized laser beam at $1064$~nm with $3$~W of the total optical power is used to trap the magnetic particle inside the hollow core of the fiber. The mechanical quadratures of the particle can be detected by  collecting the scattered light from the particle coming out of the fiber cladding as shown in the inset of Fig.~\ref{Fig1AppendixC} and focusing it onto a quadrature position sensor. Under ambient conditions of pressure, the particle's motion is heavily damped due to collisions with the background gas. However, the system can be put in the underdamped oscillating regime once the fiber is evacuated using a vacuum chamber. In our experimental setup, at $1$ mbar,  underdamped motion of the particle is confirmed by detecting a Lorentzian spectrum of motion along $y$-axis, as shown in Fig.~\ref{Fig2AppendixC}. The measured spectrum provides a resonance peak at a frequency $\sim 2 \pi\times 34.5$~kHz and damping rate $\sim$ $2\pi\times0.3$ kHz. The noise floor in the spectrum has a value $17$ pm/$\sqrt{\mathrm{Hz}}$ and the additional spikes in the spectrum arise due to the electronic noise from the detectors and data collection.
	
	%%%%%%%%%%%%%%%%%%%%%%%%%%%%%%%%%%%%%%%%%%
	%%%%%%%%%%%%%%%%%%%%%%%%%%%%%%%%%%%%%%%%%%
	\begin{figure}[t]
		\includegraphics[width=0.5\columnwidth]{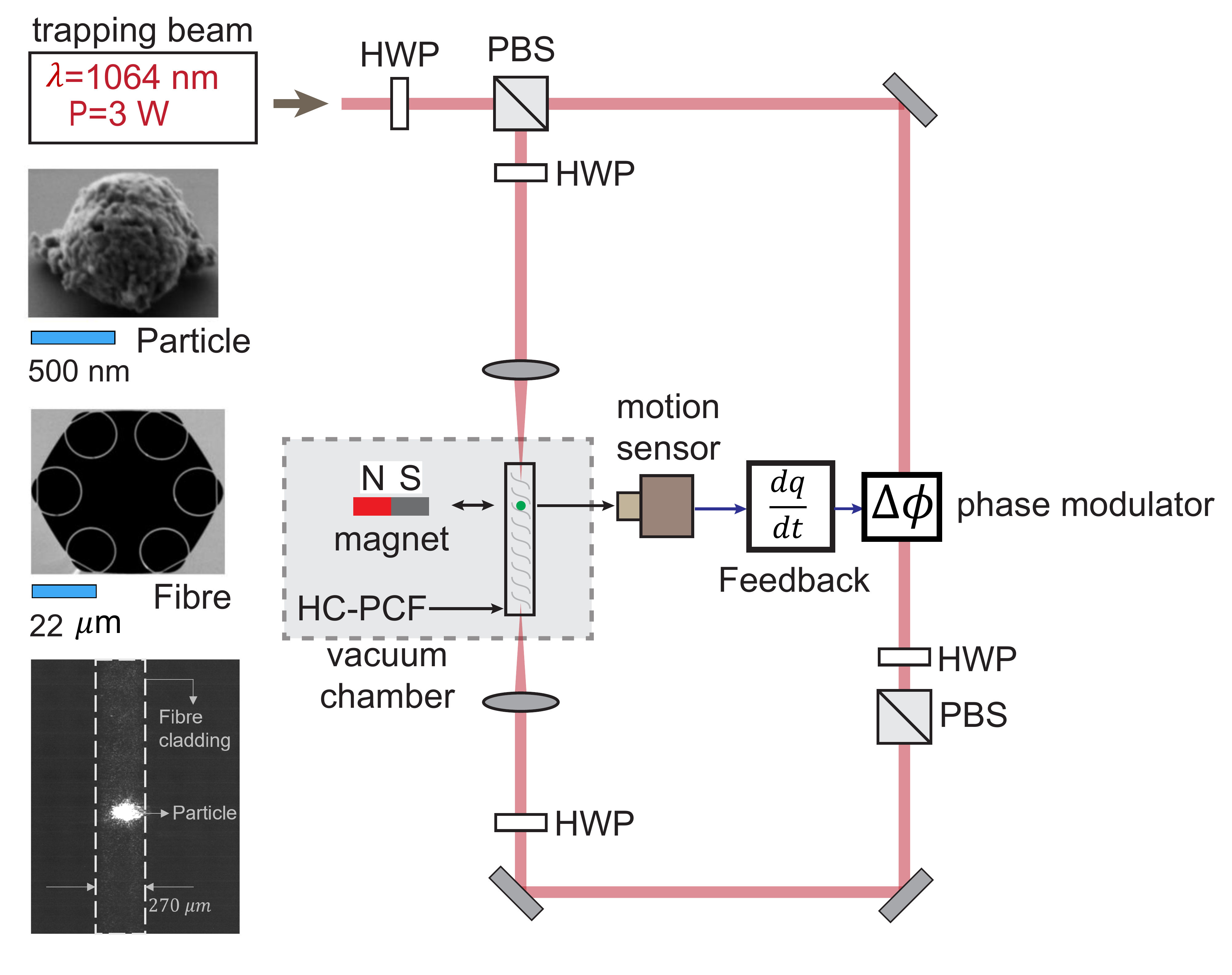}
		\caption {Schematic of the experimental setup. HWP: half-wave plate, PBS: polarizing beam splitter. The fiber is placed inside a vacuum chamber capable of reaching ultra-high vacuum. A permanent magnet is mounted on a translational stage and can be moved back and forth radially relative to the fiber in order to adjust the amplitude of the magnetic field applied to the trapped particle. Insets: Scanning electron micrographs showing the cross-section of the HC-PCF with 44 $\mu$m core diameter, the 1 $\mu$m magnetic microparticle, and snapshot of an optically trapped magnetic particle inside the core of a HC-PCF captured with a high-speed camera. The motion sensor is a quadrant photodiode which detects the position quadrature of the levitated particle. }
		\label{Fig1AppendixC}
	\end{figure}
	%%%%%%%%%%%%%%%%%%%%%%%%%%%%%%%%%%%%%%%%%%
	%%%%%%%%%%%%%%%%%%%%%%%%%%%%%%%%%%%%%%%%%%
	
	%%%%%%%%%%%%%%%%%%%%%%%%%%%%%%%%%%%%%%%%%%
	%%%%%%%%%%%%%%%%%%%%%%%%%%%%%%%%%%%%%%%%%%
	\begin{figure}[ht!]
		\includegraphics[width=0.4\columnwidth]{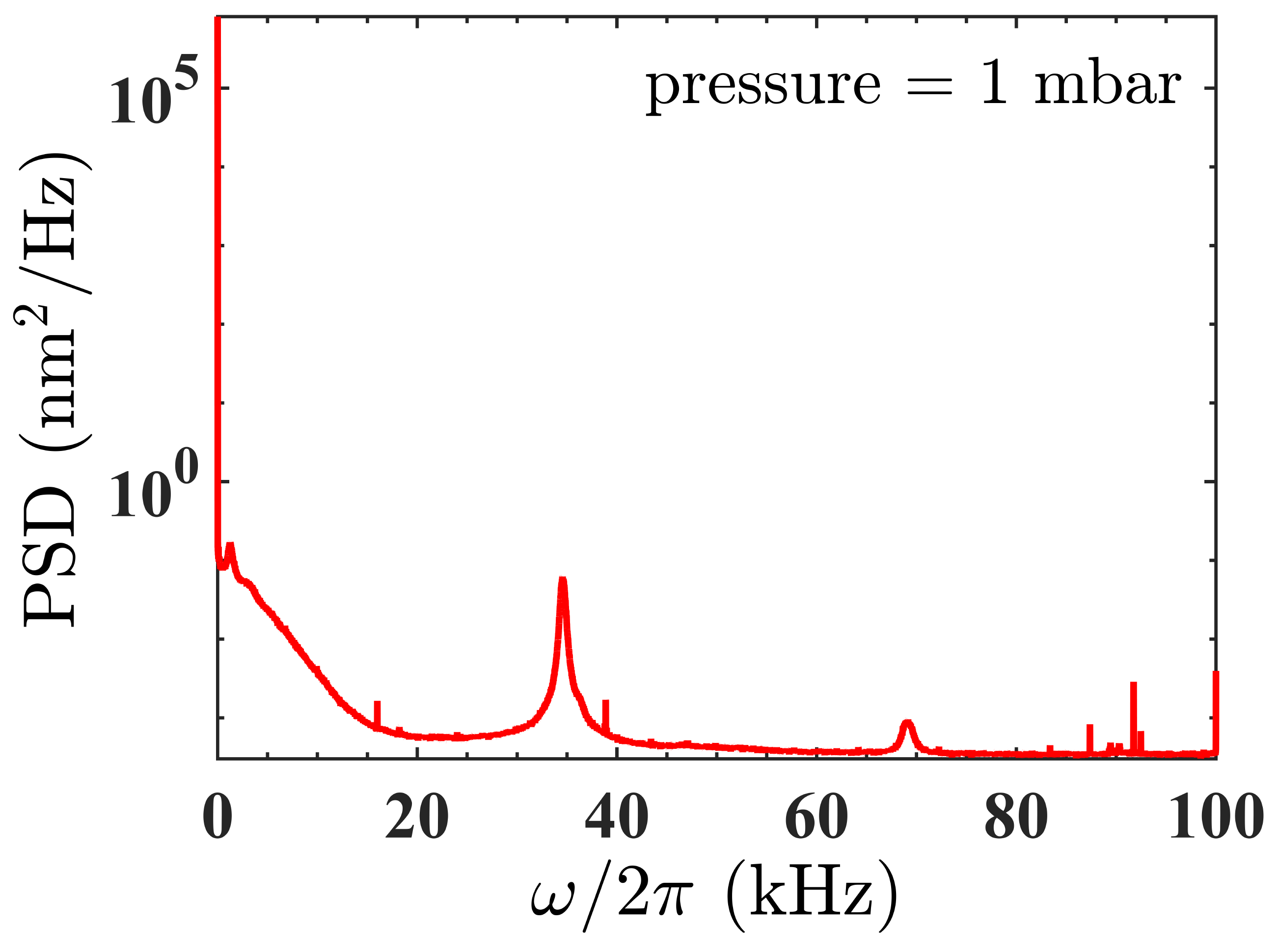}
		\caption { Experimental spectrum of the damped mechanical motion of the bound magnetic particle at a pressure of 1 mbar. The resonance peak at $\sim 2\pi\times$34.5 kHz corresponds to the underdamped motion of the particle. The additional spikes shown in the spectrum are electronic noise from our detectors and data collection. The laser power is 3 W.}
		\label{Fig2AppendixC}
	\end{figure}
	%%%%%%%%%%%%%%%%%%%%%%%%%%%%%%%%%%%%%%%%%%
	%%%%%%%%%%%%%%%%%%%%%%%%%%%%%%%%%%%%%%%%%%
	Let us remark at the possible experimental strategy to implement such a phase adaptive feedback cooling mechanism, for the setup described in Fig.~\ref{Fig1AppendixC}. The magnetic particle is trapped inside a hollow core fiber by using a linearly polarized laser beam of power $3$ W and wavelength $1064$~nm at an initial pressure of $10$~mbar. Due to collinear polarisation states of the tweezer arms, a standing wave pattern is formed inside the fiber. The starting point of the phase-adaptive feedback cooling is to detect the time varying position quadrature of the particle. This can be accomplished by imaging the particle by collecting the scattered light coming out of the fiber cladding [see inset of Fig.~\ref{Fig1AppendixC}]. The collected light is then focused onto a quadrant position sensor from which a signal  would be filtered out by a bandpass filter centered around $2\pi\times 34.5$ kHz, followed by a derivative circuit, a variable gain amplifier and a phase shifter. The output signal goes to the phase-modulator such as a Pockels cell. The Pockels cell will modulate the phase of the trapping beam in one of the tweezer arms in order to shift the standing wave pattern proportional to the instantaneous velocity of the levitated particle inside the HC-PCF.\\

	\subsection{Mie scattering in axial feedback cooling}
	\label{AppendixC1}
	When the particle diameter becomes comparable to the trapping beam wavelength \cite{pflanzerPRA2012}, it becomes essential to consider Mie scattering rather than relying solely on the Rayleigh approximation, which assumes a point-like dipole interaction. In the experimental setup described above, the magnetic particles have diameters of approximately 1–1.2 microns, causing them to span two adjacent antinodes of the standing wave trap (SWT). Under these conditions and depending on the refractive index of the material, the particle may experience equal attractive forces from each antinode and behave as a low-field seeker \cite{zemanekJOSA2002}, with its center of mass naturally aligning to the node of the SWT. Previous studies on optical conveyor belts \cite{cizmarAPL2005,cizmarPRB2006,cizmarSPIE2004} have demonstrated that Mie scatterers can be effectively trapped and transported by moving the interference fringes.

	The sign of the axial force originating from the SWT depends on whether the particle is a high-field or low-field seeker—a characteristic that is sensitive to the particle’s diameter and refractive index. Our analytical model is predicated on the existence of a restoring force acting on the particle's center of mass due to the interference fringes in the axial direction, which gives rise to an axial eigenmode. Provided that the particle’s position and momentum quadratures are measurable, the center of mass of the axial eigenmode can be cooled.

	As demonstrated above, the Mie scatterer in our experiment exhibits a well-defined axial eigenmode (see Fig. \ref{Fig2AppendixC}), thus confirming the feasibility of cooling its center-of-mass motion. For simplicity, the numerical model assumes a dipole-like gradient force on the particle's center of mass. At low pressure, when the motion of the trapped particle becomes underdamped, its center of mass remains confined within a single interference fringe. This has been verified by measuring a noise-driven axial displacement of approximately 10 nm at 1 mbar pressure, compared to a fringe width of 532 nm. Consequently, by considering the center of mass as effectively residing at the node (or antinode) of the fringe, the restoring force can be modeled as being proportional to the gradient of the standing-wave intensity, thereby justifying the numerical simulation of center-of-mass cooling.

Here we present a simulation \cite{nieminenJoA2007} of the axial optical force acting on a Mie scatterer. In our setup, a 1.2 $\mu$m magnetic particle with a refractive index of 2.11 is confined in the $y$-direction by the interference fringes of a SWT generated inside a HC-PCF using 0th-order non-diffracting Bessel beams at a wavelength of 1064 nm. Fig. \ref{FigAppendixC1} illustrates the restoring force on the particle’s center of mass for small displacements from the trap center.
	%%%%%%%%%%%%%%%%%%%%%%%%%%%%%%%%%%%%%%%%%%%%%%%%%%%%%%%%%%%%%%%%%%%%%%
	%%%%%%%%%%%%%%%%%%%%%%%%%%%%%%%%%%%%%%%%%%%%%%%%%%%%%%%%%%%%%%%%%%%%%%
	\begin{figure}[t]
		\includegraphics[width=0.4\columnwidth]{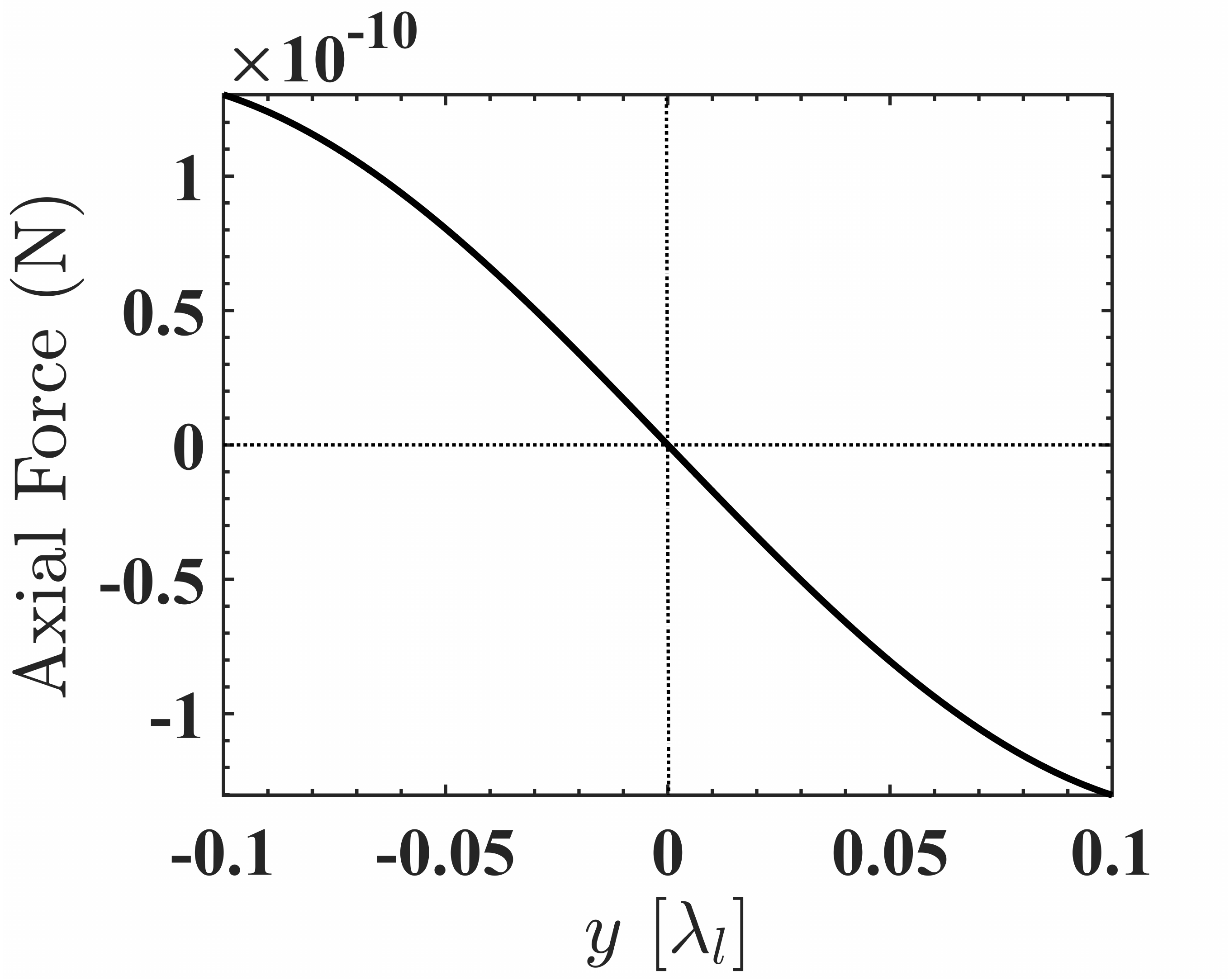}
		\caption {Axial force acting on the centre of mass of a 1.2-micron size magnetic sphere trapped by standing wave inside the hollow-core of a photonic crystal fiber, with 1 W of trapping beam power at 1064 nm and considering small displacement from the center of the fringe. The refractive index of the material is taken to be 2.11 at 1064 nm. The horizontal axis is normalized to wavelength of the trapping beam in the air ($\lambda_l=1064$ nm).}
		\label{FigAppendixC1}
	\end{figure}
	%%%%%%%%%%%%%%%%%%%%%%%%%%%%%%%%%%%%%%%%%%%%%%%%%%%%%%%%%%%%%%%%%%%%%%
	%%%%%%%%%%%%%%%%%%%%%%%%%%%%%%%%%%%%%%%%%%%%%%%%%%%%%%%%%%%%%%%%%%%%%%

\section{Estimation of imaging factor ($\eta$)}
\label{AppendixD}
Our phase-adaptive mechanism is based on the detection of particle's position which can be deduced as $y_{\mathrm{det}}(t)=y(t)+y_{zpm}\eta W^{det}(t)$. Here the estimated position is written as a sum of the true position $y(t)$ and detection noise which is modelled in terms of Wiener process. To find out the detection noise amplitude $\eta$, we write the position power spectral density $S_{\mathrm{det}}(\omega)=S^{th}_{y}(\omega)+y_{zpm}^{2}\eta^{2}S_{W}(\omega)$. Here, the first term is the thermal noise contribution written as:
\begin{align}
	S_{y}(\omega)&=\frac{2k_{B}T_{\mathrm{th}}\gamma/m}{(\omega^{2}-\Omega^{2})^{2}+\gamma^{2}\omega^{2}}\;,\label{ThermalSpectrum}
\end{align}
whereas the second term in $S_{\mathrm{det}}$ represents a contribution from the detection noise that sets the noise floor in the measurement. Note that the various terms involved in Eq. (\ref{ThermalSpectrum}) are already defined in Sec. \ref{sec2}. Now, to obtain contribution of detection noise, we write the finite Fourier transform of the Wiener process over a measurement time window $T_{w}$ as follows
\begin{align}
	\tilde{W}^{det}(\omega,T_{w})&=\int_{0}^{T_{w}}W^{det}(t)e^{-i\omega t}dt\;.\label{eqE1}
\end{align}
The sampled spectral density of the Wiener process can then be estimated as
\begin{align}
	S_{W}(\omega)&=\frac{1}{T_{w}}\langle(\tilde{W}^{\mathrm{det}})^{\ast}(\omega,T_{w})\tilde{W}^{\mathrm{det}}(\omega,T_{w})\rangle\;.\label{eqE2}
\end{align}
Using Eq. (\ref{eqE1}) in Eq. (\ref{eqE2}), we can write
\begin{align}
	S_{W}(\omega)&=\frac{1}{T_{w}}\int_{0}^{T_{w}}\int_{0}^{T_{w}}\langle W^{\mathrm{det}}(t')W^{\mathrm{det}}(t'')\rangle e^{-i\omega(t''-t')}dt''dt'\;.
\end{align}
Changing the region of integration in above equation, we get
\begin{align}
	S_{W}(\omega)&=\frac{1}{T_{w}}\int_{0}^{T_{w}}\int_{0}^{t'}\langle W^{\mathrm{det}}(t')W^{\mathrm{det}}(t'')\rangle e^{-i\omega(t''-t')}dt''dt'+\frac{1}{T_{w}}\int_{0}^{T_{w}}\int_{t'}^{T_{w}}\langle W^{\mathrm{det}}(t')W^{\mathrm{det}}(t'')\rangle e^{-i\omega(t''-t')}dt''dt'\;.
\end{align}
For a Wiener process, two time correlation function is expressed as $\langle W^{\mathrm{det}}(t')W^{\mathrm{det}}(t'')\rangle=\mathrm{min}(t',t'')$. For $0\le t''\le t'$, $\langle W^{\mathrm{det}}(t')W^{\mathrm{det}}(t'')\rangle=t''$ and for $t'\le t''\le T_{w}$, $\langle W^{\mathrm{det}}(t')W^{\mathrm{det}}(t'')\rangle=t'$. Using this and performing the integration, the spectral contribution due to detection noise in the large $T_{w}$ takes following form
\begin{align}
	S_{W}(\omega)\approx\frac{2}{\omega^{2}}\;.\label{eqE6}
\end{align}
The estimated noise floor for the detection scheme under consideration is 17 pm/$\sqrt{\mathrm{Hz}}$. Around $\omega\approx\Omega$ and using Eq. (\ref{eqE6}), we find $\eta=\frac{17\Omega}{y_{\mathrm{zpm}}\sqrt{2}}\frac{\mathrm{pm}}{\sqrt{\mathrm{Hz}}}$.

\section{Optimum value of $\mathcal{C}$}
\label{AppendixE}
To obtain the optimum value of the amplifier gain $\mathcal{C}$ for any $\tau$, we minimize the final occupancy presented in Eq. (\ref{FinalOccupancy}). This allows us to write following generalized expression for the optimized gain 
\begin{align}
	\mathcal{C}_{opt}^{\tau}&=-\frac{\gamma}{\beta B_{1}^{\tau}}+\sqrt{\frac{\gamma^{2}}{\beta^{2} {B_{1}^{\tau}}^{2}}+\frac{2\Omega^{2}\gamma n_{\mathrm{th}}}{\eta^{2}\beta^{2}{B_{2}^{\tau}}^{2}}}\;,
\end{align}
where $B_{1}^{\tau}=e^{\gamma\tau/2}\frac{\sin(\Omega\tau)}{\Omega\tau}$ and $B_{2}^{\tau}=\frac{\sin(\Omega\tau/2)}{\Omega\tau/2}$.

\section{Numerical procedure}
\label{AppendixF}
Now we describe the numerical procedure to obtain the mechanical oscillator dynamics in the presence of thermal and measurement noise. For this, we write the following coupled difference equations
\begin{subequations}
	\begin{align}
		dq &= \Omega pdt\;,\label{Eq1MotionNoise}\\
		dp &=-\gamma pdt-\Omega qdt-\frac{\beta\mathcal{C}}{\Omega} \Bigg(\frac{q(t)-q(t-\tau)}{\tau}\Bigg) dt+\sqrt{2\gamma n_{\mathrm{th}}}dW^{\text{th}}(t)-\mathcal{A}\sum_{m=1}^{n_{\tau}}dW_{n-n_{\tau}+m}^{\mathrm{det}}\label{Eq2MotionNoise}\;,
	\end{align}
\end{subequations}
where the action of the thermal bath is included as a Wiener process $dW^{\mathrm{th}}(t)$ of zero average and variance equal to numerical time $dt$. Numerically, we can express the Wiener increments in terms of a Gaussian distribution \cite{jacobs2010book} as $dW^{\mathrm{th}}(t)=\sqrt{dt}\mathcal{N}(0,1)$, where $\mathcal{N}(0,1)$ represents a normal distribution of unit variance from which a random variable is to be drawn. On the other hand, the last term in Eq. \eqref{Eq2MotionNoise} represents Wiener process entering during the measurement process and is written as a discrete sum over the Wiener increments with increment of time $dt$. Numerically, we model it as $\sum_{m=1}^{n_{\tau}}dW^{\text{det}}_{n-n_{\tau}+m}=\sqrt{\tau}\sum_{m=1}^{n_{\tau}}\mathcal{N}_{m}$, where $\mathcal{N}_{m}$ is a random number drawn from a normal distribution at each time evolution step. To numerically solve Eqs. (\ref{Eq1MotionNoise}-\ref{Eq2MotionNoise}), we use the Runge-Kutta fourth-order (RK4) method with a time step $dt=5\times 10^{-3}/\Omega$ that ensure the numerical stability.\\
\begin{figure}[t]
	\includegraphics[width=0.8\columnwidth]{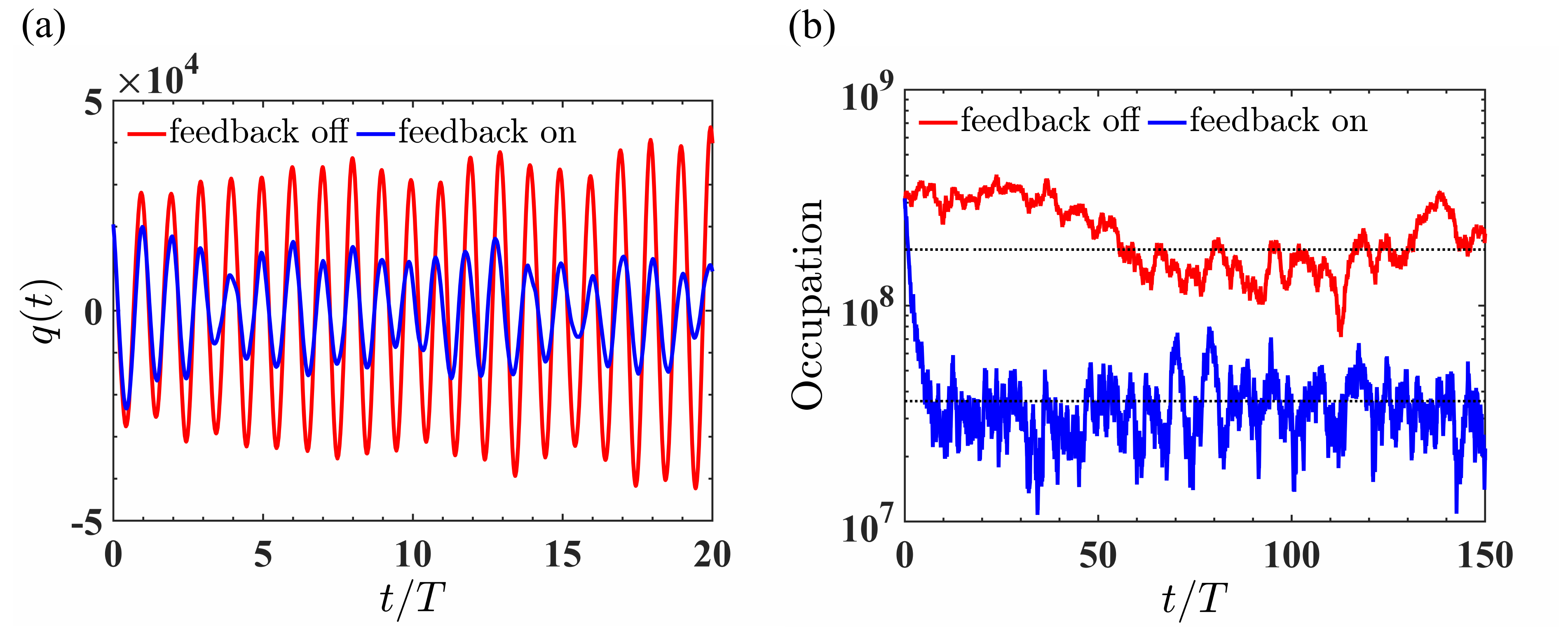}
	\caption {Temporal evolution of the phonon occupancy in the absence  (solid red line) and presence (solid blue line) of cooling scheme. The phonon occupancy is obtained  from the numerical simulation of Eqs. (\ref{Eq1MotionNoise}-\ref{Eq2MotionNoise}). The dotted lines show the validity of the analytically predicted final occupancy given by Eq. (\ref{FinalOccupancy}). Here $p=1$ mbar and rest of the parameters are same as in Fig. \ref{fig4}.}
	\label{figAppendixF}
\end{figure}
Using above numerical procedure, we simulate the difference equations (\ref{Eq1MotionNoise}-\ref{Eq2MotionNoise}) for an oscillator undergoing thermalization with a thermal environment of effective occupancy $n_{\mathrm{th}}=1.8\times 10^{8}$ and at a pressure of 1 mbar.  Fig. \ref{figAppendixF}(a) shows the resulting position dynamics while Fig. \ref{figAppendixF}(b) gives the phonon occupancy.   These results are plotted both in the absence and presence of feedback cooling scheme. As shown in Fig. \ref{figAppendixF}, in the absence of the proposed feedback scheme, the system quickly achieves equilibrium with its thermal environment thereby attaining a high thermal  phonon-occupancy governed by $\bar{n}=n_{\mathrm{th}}$. However, when the phase-adaptive feedback cooling is switched on for $\tau/T=7.9\times 10^{-4}$, the initial high phonon occupancy decays and eventually it saturates at a final value $\bar{n}$ governed by our analytical expression in Eq.~ (\ref{FinalOccupancy}).

\section{Laser frequency noise}
\label{AppendixG}
The instantaneous frequency fluctuations of a trapping laser originate from its randomly varying phase. To model it, we express the amplitude of electric field of a single-mode electromagnetic field as \cite{rablPRA2009}
\begin{align}
	\mathcal{E}_{0}(t)=|\mathcal{E}_{0}|e^{i\phi_{l}(t)}\;, \label{Eq1AppendixG}
\end{align}
where $\phi_{l}(t)$ is a Gaussian random variable with zero mean, $\langle\phi_{l}(t)\rangle=0$ and variance $\langle\phi_{l}^{2}(t)\rangle=\sigma_{\phi_{l}}$. This phase noise affects the amplitude of the counter-propagating fields, so that the total electric field forming the standing-wave pattern inside the HCPCF is given by
\begin{align}
	\mathcal{E}(y,t)&=2|\mathcal{E}_{0}|e^{i\phi_{l}(t)}\cos[k_\ell y+\phi(t)]\;.\label{Eq2AppendixG}
\end{align}
By averaging this field over many realizations and using the Gaussian noise property,  $\langle e^{i\phi_{l}(t)}\rangle= e^{-\sigma_{\phi_{l}}/2}$, we obtain 
\begin{align}
	\langle\mathcal{E}(y,t)\rangle&=2|\mathcal{E}_{0}|\langle e^{i\phi_{l}(t)}\rangle\cos[k_\ell y+\phi(t)]=2|\mathcal{E}_{0}|e^{-\sigma_{\phi_{l}}/2}\cos[k_\ell y+\phi(t)]\;.\label{Eq3AppendixG}
\end{align}
Following the procedure described in Sec. \ref{sec2}, we find that the laser phase noise rescales the trapping frequency from $\Omega$ to $\Omega e^{-\sigma_{\phi_{l}}/2}$. Consequently, the parameter $\beta$ is modified to $\beta e^{-\sigma_{\phi_{l}}}$, which leads to a corresponding change in the effective cooling rate from $\bar{\Gamma}$ to $\bar{\Gamma}e^{-\sigma_{\phi_{l}}}$. Within this framework, the detection noise amplitude $\eta$ and optimized gain factor $\mathcal{C}_{opt}$ are considered independent of the phase noise.

\end{widetext}
\end{document}